\documentclass[twocolumn, showpacs,preprintnumbers,amsmath,amssymb]{revtex4}
\usepackage{graphicx} % use pdf for arXiv
\usepackage{dcolumn} % single column for PRF, 2-column for PRE
\usepackage{bm}

\newcommand{\be}{\begin{equation}}
\newcommand{\ee}{\end{equation}}

\begin{document}

\title{Strain stiffening due to stretching of entangled particles in random packings of granular materials}

\author{Eric Brown}
\email{ericmichealbrown@gmail.com}
\affiliation{Eric Brown Labs LLC, New Haven, CT 06511}
\affiliation{Department of Physics, University of California, Merced, CA 95343}
\affiliation{James Franck Institute and Department of Physics, The University of Chicago, Chicago, IL 60637}
%\affiliation{Department of Physics, Southern Connecticut State University, New Haven, CT 06515}

%\author{Sulimon Sattari}
%\author{David Brantley}
%\affiliation{School of Natural Sciences, University of California, Merced, CA 95343}
%\author{Alice Nasto}
%\affiliation{Department of Mechanical Engineering, Massachusetts Institute of Technology, Cambridge MA 02139}
 %\author{Xuan-Truc Nguyen}
 %\author{Henos Musie}
 %\author{Tammer Abiyu}
 %\affiliation{Department of Mechanical Engineering and Materials Science, Yale University, New Haven, CT 06511, USA}
 \author{Kevin A. Mitchell}
\affiliation{Department of Physics, University of California, Merced, CA 95343}
\author{Alice Nasto, Athanasios Athanassiadis,  Heinrich M. Jaeger}
\affiliation{James Franck Institute and Department of Physics, The University of Chicago, Chicago, IL 60637}
%Eric Brown1,2,3,  Alice Nasto3, Athanasios Athanassiadis3, Heinrich Jaeger3, Sulimon Sattari1, David Brantley1, Kevin Mitchell1,  Xuan-Truc Nguyen2, Henos Musie2, Tammer Abiyu2, 1University of California, Merced, 2Yale University, 3University of Chicago

\date{\today}

\begin{abstract} 

Stress-strain relations  for random packings of entangling chains under triaxial compression can exhibit strain stiffening and sustain stresses several orders-of-magnitude beyond typical granular materials.  X-ray tomography reveals the transition to this strong strain stiffening occurs when chains are long enough to entangle an average of about one chain each, which results in system-filling clusters of entangled chains, similar to the Erd{\"o}s-R{\'e}nyi model for randomly connected graphs.  The number of entanglements is nearly proportional to the area surrounded by entangling particles with an excluded volume effect, thus the existence of system-filling clusters of entanglements can be predicted  assuming random particle positions and orientations with an excluded volume effect if the particle shapes in the packing are known.  A tendency was found for chain links to stretch when the packing was strained
 This suggests that the strength of these packings comes from stretching of the links of chains, but only when the system-filling network of entanglements provides constraints that prevents failure by shear banding, so that particles must be deformed to move further under strain.   The slope of the stress-strain relation of a packing  can be calculated from a mean-field model consisting of the product of the effective extensional modulus of the chain, packing fraction, probability of stretched links, and the ratio of strain of stretched links to packing strain.  In this model,  the increasing slope of the stress-strain curve is mainly due to the fraction of stretched links increasing with strain, and assuming the fraction of stretched links is proportional to strain results in a quadratic prediction for the stress-strain curve.  The stress-strain model requires as input measurements of the ratio between local particle deformation and global average strain, and the probability of stretching for non-rigid particles, resulting in a quadratic curvature that agrees with experiments within the run-to-run variation ($30\%$).        This model for the stress-strain relation is shown to be generalizable to different shapes of entangling particles by applying it to staples, where the packing strength comes from the bending of staples instead of stretching links.  The permanent plastic deformation of staples allows measuring statistical quantities from inspection of a poured out sample after a triaxial compression, without the need for in-situ imaging.   Both the probability of staples bending and the average bend angle of the arms were found to increase with strain, and these inputs into the model result in a quadratic curvature of the stress-strain that agrees with experiments within the model uncertainties ($37\%$).
\end{abstract}

%physSH: granular solids, granular chains, tomography, granular packing
%62.20.mm structural failure:fracture
%81.70.Tx Computed tomography
%83.80.Fg rheology of Granular solids

\maketitle

\section{Introduction}
%problem statement
%Most random packings of granular materials have an initially linear stress-strain relation, which levels off as they reach a maximum stress, which is typically limited by attractions between particles or an external confining pressure \cite{LW69}.  On the other hand, random packings of granular chains exhibit strong strain stiffening, whereby the ratio of stress to strain increases as the material is further strained, and they can sustain stresses that exceed the contributions from attractions and external confinement by several orders of magnitude \cite{BNAJ12}.  %This strong strain stiffening was found only if the chains were long enough to form system-filling clusters of entangled chains \cite{BNAJ12}.  
%Strain stiffening can also be seen in other systems of particles that can entangle \cite{MRCSJ16, MJ12, MFT15}.  
 %In this paper, we follow up on a previous letter \cite{BNAJ12} to present and test quantitative, low-dimensional, and generalizable models to predict the particle shapes that can produce strong strain-stiffening, and describe the resulting stress-strain relation.

%motivation
The stress-strain relations of randomly packed granular materials have long been studied because of the importance of soil mechanics to the foundations of buildings.  %, but is usually limited to spherical or otherwise convex particles that do not entangle
Random packings of granular materials of nearly-spherical natural shapes typically have an initially linear stress-strain relation, which levels off to an approximately  constant stress at large strain.   The limiting stress is typically determined by the stronger of either attractions between particles or an external confining pressure, rather than the particle modulus, because particles in packings can rearrange under shear in response to stress \cite{LW69}.  
On the other hand, random packings of granular materials that entangle or interlock can exhibit strong strain stiffening, whereby the ratio of stress to strain increases as the material is further strained, and they can sustain stresses that exceed the contributions from attractions and external confinement by several orders of magnitude \cite{BNAJ12, DHRSPD18, MRCSJ16, MJ12}.  The source of this strength and why it results in strain stiffening remains unknown.

Studying the mechanics of randomly packed entangling granular materials may lead to understanding of the mechanics of a broader class of materials that can entangle.  This includes random packings of rigid particles with concave regions which can interlock \cite{KFL09, MJ12}.  Such particles have been of interest, for example as a pourable construction material with high strength but without need for a binding material \cite{Danel53,DM16}.  Strain stiffening is observed in athermal gels formed by semiflexible fibers, where it is interpreted as a critical scaling behavior beyond a critical strain where fiber stretching becomes the dominant source of stress \cite{OVLSGMJ16}.  There are many strain stiffening systems consisting of entangled systems of flexible fibers, including collections of disorganized matted fibers like cotton balls  \cite{PVA05}, some knitted textiles \cite{WSTQLYKD18},
 %not clear why stiffening. due to tightening of loops?  Fabric compressed in middle by large amount during test, might produce some nonlinearity
and birds nests \cite{WBGK20}, where the mechanics responsible for strain stiffening are not yet understood.  There are opportunities for unique control of mechanical properties in entangled systems. For example, entangled structures of hard particles can have bending stiffness that is tunable with applied pressure over a much higher range than traditional soils \cite{WLHAD21}.   
The ability to manufacture particles with arms that can be activated to control entanglements opens up more possibilities for controllable mechanics  \cite{STBG23}.
There have been several observations that suggest increased structural support due to entanglement of granular materials.   Random packings of entangling particles have been found to form tall free-standing structures in experiments with U-shaped particles (i.e.~staples) \cite{GFHG12}, Z-shaped particles \cite{MRCSJ16},  and  star-shaped particles \cite{DM16}, and in simulations of S-shaped and U-shaped particles \cite{KMMRA22}.  Such entangled structures can  remain relatively long-lived even under vibration  \cite{GFHG12, Franklin12}. % and  relax more slowly compared to packings of sphere-like particles \cite{ZCRJN09}.   
Free-standing structures of particles demonstrate that such packings have some strength that exceeds their weight, without the need for any confining stress or attractions, which is usually what limits the strength of packings of convex or nearly spherical particles.  

%strength measurements
The strength of soils and other granular packings is typically measured in triaxial compression experiments, where a cylindrical material is compressed along one axis with a controlled pressure applied at the other boundaries.  Strain stiffening with stresses well above the confining stress has been reported in random packings of chains \cite{BNAJ12, DHRSPD18}, Z-shaped particles \cite{MRCSJ16}, U-shaped particles \cite{MRCSJ16}, and other non-convex particles \cite{MJ12}, as well as two-dimensional simulations of packings of soft U-shaped staples \cite{MFT15}, and three-dimensional simulations of packings of chains \cite{SH22}.  Long rod-shaped particles can form tall free-standing structures as well \cite{Philipse96}, and exhibit strain stiffening \cite{BWBKGK22}, although when unbent they do not clearly entangle in the same was as non-convex shapes.  It has been suggested that long rods bend under stress enough to produce concave regions that can entangle, and  as the number of contact points between particles increase with strain, the distance between contact points on the rods decreases, which could increase the bending stiffness with strain to cause strain stiffening \cite{BWBKGK22}.

%polymer models
%Strain-stiffeneing in polymers has been attributed  However, those arguments only explained much weaker strain stiffening in which the stress increase is less than an order of magnitude \cite{VLM09, HO10}, so no additional confining stress mechanism was necessary in those cases.  strain stiffening is known to occur in many polymeric materials as well \cite{Treolar49}.  Theories suggest strain stiffening could depend on many factors in polymers including  chain stiffness, density, temperature, strain rate, and in particular on structures such as entanglements between different chains \cite{BJ95, HR06, VLM09, HO10}.   It is expected that entanglements can lead to strain stiffening and an increased transient strength in polymers, but over time the entanglements unentangle due to thermal motion.  In  athermal, quasistatically deformed granular chain packings, temperature and strain rate are not relevant parameters, and entangled granular systems may be analogous to the high shear rate, low temperature limit in polymers where entanglement do not have time to relax thermally.   

%polymer
Strain stiffening is known to occur in many thermal polymeric materials as well, where entanglements can be spontaneously driven by thermal motion \cite{Treolar49}.  Models for strain stiffening in polymers usually rely on the thermodynamics of wormlike chains, and have been applied only to relatively weak strain stiffening in polymer systems  \cite{BJ95, HR06, VLM09, HO10}.  The wormlike chain model has been adapted to athermal granular systems with localized strain using the mechanism of self-amplified friction  \cite{DHRSPD18}, but we find it does not characterize the type of strain stiffening we have observed in triaxial compression tests of entangling granular materials (see Appendix \ref{sec:expmodel}).

%...Monte Carlo simulation of excluded areas which can lead to calculation of entanglement probabilities have been made for super-ellipse particles (including for example semicircular arcs and staples) \cite{KBF21}.  However, these were done in two dimensions only.

%Previous work on granular chains focused on their packing structure near the jamming transition, but did not address their response to stress \cite{KFL09, ZCRJN09, LRR11}.  We go beyond those works by first demonstrating that packings of granular chains of sufficient length exhibit strain stiffening and can sustain stresses orders-of-magnitude greater than those of unlinked granular materials.  We then use x-ray tomography to measure the precise packing structure and identify entanglements to demonstrate quantitative connections between entanglements and strain stiffening. 

% in analogy to polymers. The free volume in the packing was found to increase with chain length and level off when the chain length greatly exceeded the persistence length, analogous to the behavior of the glass transition temperature for polymers

%compression vs extension
 A two-dimensional simulation of granular chains exhibited weak strain stiffening under compaction in a hard-walled box, and it was suggested that compression of those loops could be responsible for strain stiffening \cite{LRR11}.  In two dimensions, the chains form loops (but cannot entangle).   As the loops are compressed, they become stiffer, leading to an overall strain stiffening.  It is not yet clear if this compression can lead to strain stiffening in three dimensions, where the constraints provided by loops do not necessarily lead to compression of the loop.  %An important difference between triaxial compression experiments and hard-wall simulations is that in triaxial compression experiments, the side walls are compliant compared to the rigid side walls in the simulation.  Rigid side walls can easily sustain the transmission of applied stress through the packing, but the compliant side walls can only support a limited stress.   
%  A different mechanism is likely required to explain the experimental observation of supported stresses that exceed the confining stress by orders of magnitude \cite{BNAJ12}.  
On the other hand, simulations of random packings of non-convex particles reveal tensile stresses between links of entangled chains \cite{KMMRA22}. It remains to be seen what role tension on links plays in strain stiffening.
%We note that we observed an overall dilation rather than compaction during several of our experiments that exhibited strong strain stiffening (see Fig.~\ref{fig:phi} in the appendices).  

%block shear band vs. entanglement
 It has been shown in 2-dimensional simulations of sheared granular piles that even a single chain can increase strength and produce strain stiffening by suppressing the usual failure mode of 
 shear banding when the chain is placed so that it crosses the path of the shear band \cite{RR13}.    Suppression of shear banding is likely necessary for the strength to greatly exceed the external confining stress.  This may be similar to the way that long fibers interspersed in random packings can suppress the formation of shear bands and increase strength in composite materials such as concrete, but typically the strength increase is not more than a factor of 2 \cite{AHFW92}.   A stronger constraint seems to be required to explain the observation of stresses several orders of magnitude times the confining stress \cite{BNAJ12}.  %Since we observed that strong strain stiffening coincided with packings that had system-filling clusters of entanglements  \cite{BNAJ12}, we propose that entanglements can provide an effective confining stress by creating internal constraints that result in stretching of the chains and much larger supported stresses. 
 %and constraints from the links contribute to rigidity \cite{ZCRJN09}.  

%follow up -- constratints -> strength
It has been argued that constraints from inter-particle entanglement allow a stronger structure in U-shaped staples \cite{GFHG12,BNAJ12}.  In our previous letter, we demonstrated that random packings of granular chains can exhibit strong strain stiffening in triaxial compression experiments, and they can sustain stresses that exceed the contributions from attractions and external confinement by 3 orders of magnitude \cite{BNAJ12}.  It was shown that strong strain stiffening occurs not just when there are a few entanglements, but when chains are long enough that on average each chain entangles at least one other chain, which is enough to result in clusters of entanglements that are system-filling (the clusters include nearly every chain in the system) \cite{BNAJ12}.  We argued that these clusters of entanglements can provide enough constraints to prevent the usual failure mode of shear banding, thus allowing the system to support much higher stresses than could be supported by the confining stress at the boundary.  However, this idea of constraints from entanglement has not yet led to models for the stress-strain relationship of packings, or a model that predicts which particle shapes will entangle to produce strong strain stiffening.  In this article, we follow up on our previous letter \cite{BNAJ12} to develop models to address both of these issues.

%summary
In the remainder of this paper we develop models to predict which particle shapes will lead to system-spanning clusters of entanglements and strong strain stiffening in random packings, and the corresponding stress-strain curve.  Section \ref{sec:methods} explains the methods for stress-strain measurements.  Section \ref{sec:stressstrain} reviews stress-strain curves from triaxial compression measurements on random packings of granular chains from our previous letter \cite{BNAJ12}, which will be used to motivate and test quantitative models.  Section \ref{sec:area} shows measurements of the area enclosed by chains from x-ray tomography of random packings of granular chains.   We use this as input to a model for  the probability of another chain passing through this area and thus being entangled in a random packing, to determine whether there will be system-filling clusters of entanglements and thus strong strain stiffening in Sec.~\ref{sec:clusters}.  Section \ref{sec:linklength} shows measurements of link stretching when the packing is under strain to determine whether strain stiffening relates to tension on links or compression of loops. In section \ref{sec:model}, we work out a quantitative prediction for the stress-strain curve based on the increased probability of chains being stretched at higher strain, and the amount of stretching.  The result is compared to stress-strain measurements for chain packings.  In Sec.~\ref{sec:staples}, we show how this model for the stress-strain relation can be applied to another type of entangled granular material: rigid U-shaped staples.  In the appendices, we include additional supporting calculations, structural measurements obtained from x-ray tomography measurements, and particle stiffness measurements.

\section{Methods}
\label{sec:methods}

\begin{figure}
\includegraphics[width=2.5in]{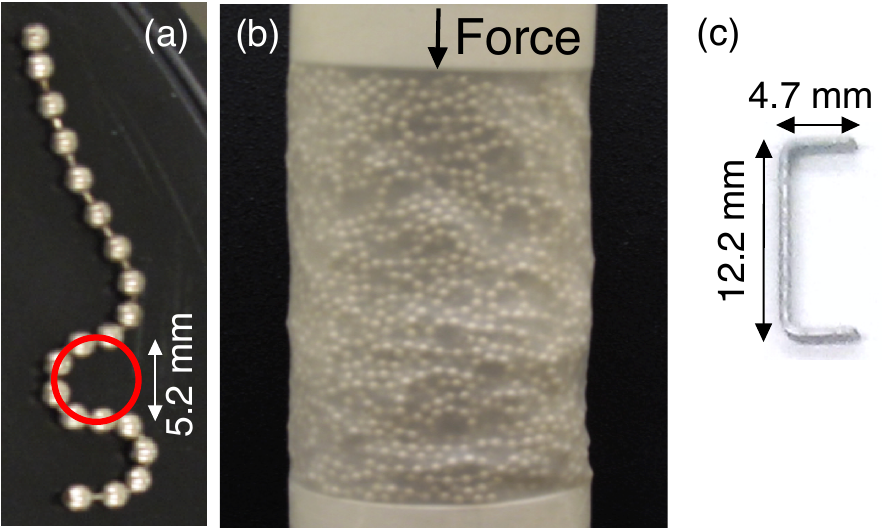}
\caption{(a)  An individual chain with length $N=20$ beads.  The minimum loop size $\xi=8$ beads without material deformation is highlighted by the red circle. (b) Cylindrical packing of a chain with length $N\approx 10^4$ inside a thin latex membrane used for triaxial compression measurements. (c) An individual staple. Chain images reproduced from \cite{BNAJ12}.
}
\label{fig:chains}
\end{figure}

As a model material that can entangle, we used macroscopic granular chains consisting of hollow spherical beads shown in Fig.~\ref{fig:chains}a.  They  are flexibly connected into chains by enclosing dog-bone-shaped brass links.  These loose connections have  zero stiffness for small extension or bend angles, but they have a maximum extension and bend angle  beyond which they do have a stiffness and particle deformation occurs.  This defines a minimum loop circumference $\xi$ the chains can bend into without material deformation, as shown in Fig.~\ref{fig:chains}a.  

We prepared packings of chains for triaxial compression experiments \cite{LW69}.   For these measurements we used brass beads with diameter $a=2.06\pm0.02$ mm (all uncertainties are presented as one standard error, corresponding to a 68\% confidence interval) and minimum loop circumference $\xi=8$ beads.   Measurements of the stiffness of individual chains in response to tension and bending are shown in Appendices \ref{sec:chainstiffness} and \ref{sec:chainbending}.    Each chain in a packing was cut to the same length $N$ in units of the number of beads. The chains were poured a few at a time under gravity into a flexible elastic membrane in a cylindrical shape with solid end caps on the top and bottom, as shown in Fig.~\ref{fig:chains}b.    The granular material was compacted by dropping the sample in the membrane attached to the bottom end cap from a height of 30 mm, for 10 repetitions.  The flexible membrane stretches just enough to provide a confining stress to hold up the weight of the sample and endcap, and allows for radial expansion when the sample is strained.  We report data for end caps with diameter 50.8 mm and initial sample heights approximately 50 mm.    The heights vary by a few mm from run to run due to different levels of compaction (packing fraction values are shown in Appendix \ref{sec:phi}).    The chains can be seen in Fig.~\ref{fig:chains}b to form tight loops within the packing, as small as the minimum loop circumference $\xi$ \cite{ZCRJN09}.    

To measure the stress-strain relation of the packing, we quasi-statically pushed the top end cap downward to compress the sample at a constant strain rate of no more than $4\times10^{-3}$ s$^{-1}$ while measuring the force on the end cap and its height.  The average compressive stress $\sigma$ is calculated as the force divided by the initial cross-sectional area (the circular cross-section area of the end cap), and the compressive strain $\gamma$ is calculated as the sample height change (positive downward) divided by the initial height.  We performed additional measurements with varying packing procedures, different strain rates, different end cap diameters, and different sample heights.  While there was some slight quantitative variation in $\sigma(\gamma)$ for these different parameters, it was comparable to the run-to-run variation of 30\% for repeated tests,  and the results did not qualitatively depend on any of these parameters, so we report only the data for the aforementioned parameter values in the following sections.

To investigate the role of packing structure in strong strain stiffening, we performed x-ray tomography on aluminum chains with bead diameter $a=2.5$ mm and minimum loop size $\xi = 7.5$ beads in cylindrical packings with diameter and height of 35 mm each.  The tomography provided the positions of each bead and link in three dimensions with a precision of 0.017 bead diameters.  
Methods from these experiments were already presented in \cite{BNAJ12}. 

Compression tests were also performed on packings of staples, using similar procedures as for chains, with samples of diameter of 38.3 mm and initial height 60 mm, compressed at a strain rate of $1.4\times10^{-3}$s$^{-1}$. The staples were galvanized steel % with bending modulus $E=9\cdot 10^{10}$ Pa.   
and U-shaped with $90^{\circ}$ bends, with arms enclosing an area $L_x=4.2$ mm by $L_y=11.3$ mm long as illustrated in Fig.~\ref{fig:chains}c.   The arms and backbone each  have width $w=1.24$ mm (into the page in Fig.~\ref{fig:chains}c, and thickness $t=0.48$ mm.   Our packings of staples had an average packing fraction $\phi=0.182$.

\section{Results}

\subsection{Strong strain stiffening}
\label{sec:stressstrain}

\begin{figure}
\includegraphics[width=3.2in]{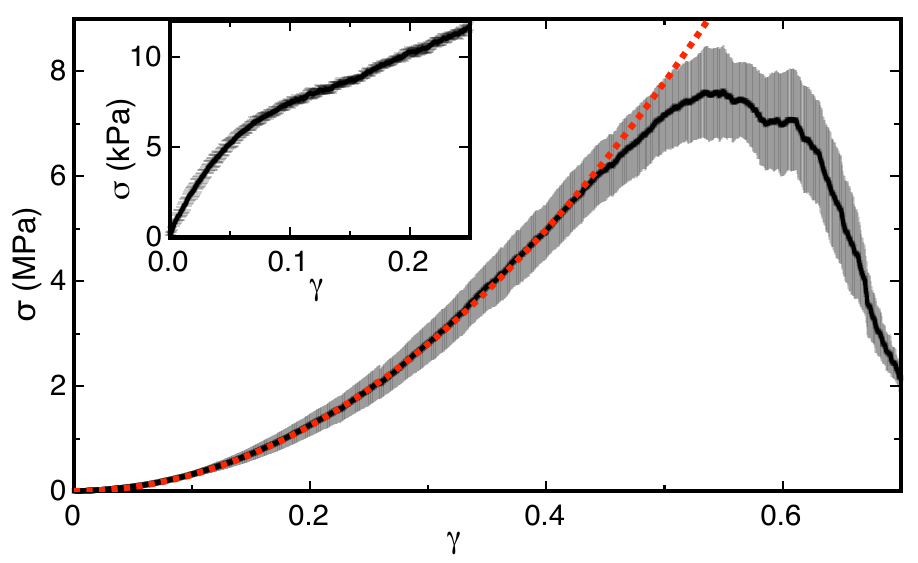}
\caption{(color online) Stress $\sigma$ vs. strain $\gamma$ for packings of chains with length $N\approx10^4$ beads.  The shaded band represents 1 standard deviation based on 5 repeated measurements. Strong strain stiffening is seen as the positive curvature in $\sigma(\gamma)$, about $10^3$ times stronger than packings of unlinked spheres ($N=1$), shown in the inset.  Data is reproduced from \cite{BNAJ12}.  Dotted red line: quadratic fit to $\sigma(\gamma)$ for $\gamma < 0.45$. % Inset:  Filled symbols: $\sigma(\gamma)$ for $N=1$ (unlinked beads). Open symbols:  reproduction of $N\approx10^4$ data for comparison on the smaller scale.
}
\label{fig:longchain}
\end{figure}

In this subsection, we summarize results from our previous letter \cite{BNAJ12}.  We show in Fig.~\ref{fig:longchain} the stress-strain relation $\sigma(\gamma)$ for a packing of chains with length $N\approx10^4$ beads,  much longer than their minimum loop circumference $\xi=8$ beads.  Strong strain stiffening can be seen as the region of positive curvature of $\sigma(\gamma)$.  A striking feature is that the stress reaches the order of 10 MPa before the packing fails, which is about $10^3$ times the stress from the weight of the packing pushing on the elastic membrane at the boundary which usually determines the scale of maximum strength of packings of convex particles.  The latter behavior is shown for $N=1$ (unlinked beads) in the inset of Fig.~\ref{fig:longchain}, where the positive slope for $\gamma \stackrel{>}{_\sim} 0.05$ matches the contribution of the membrane stiffness as it is deformed by the radially expanding packing, which is steeper while the packing is initially dilating.  In other words, the strength of unlinked bead packings is largely determined by the confining stress rather than representing an intrinsic strength of the packing.  The maximum strength of the long chain packing reaches up to the order of 1/10 the strength per density of the solid brass that makes up the chains.  During the compression of the long chains in Fig.~\ref{fig:longchain} we started to hear chains break at the point where the maximum stress was reached, and after each measurement that exceeded this ultimate stress, we counted approximately 10\% of the links in the chains to be broken.   This breaking, and the ability to support stresses much greater than the confining stress suggests that the strength of the links are the limiting factor in the packing strength.  

\begin{figure}
\includegraphics[width=3.in]{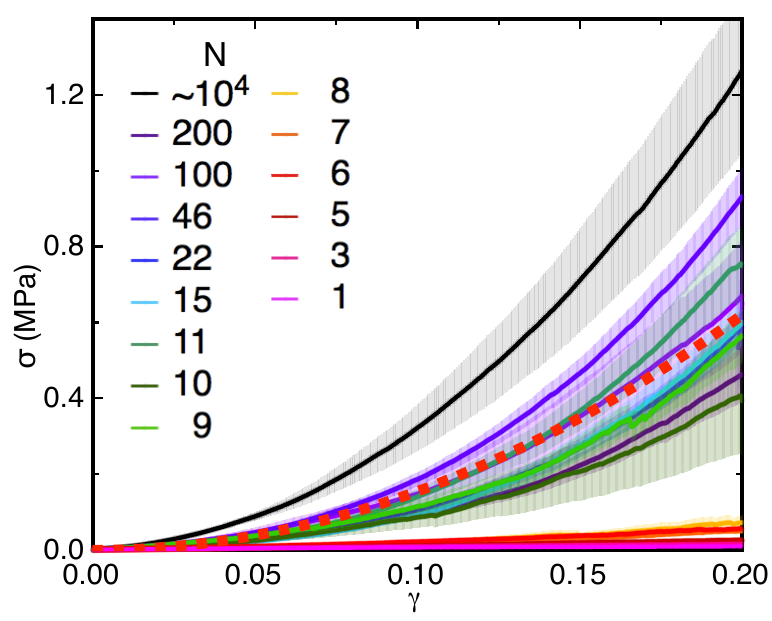}
\caption{(color online) Stress $\sigma$ vs. strain $\gamma$ for different chain lengths $N$ shown in the legend.  The curves for $N \ge 9$ (cool colors) group into a band which exhibits strong strain stiffening.  These chain lengths with strong-strain stiffening are also those which are expected to have system-filling clusters of entanglements. Data is reproduced from \cite{BNAJ12}. Dotted red line:  model prediction of $\sigma(\gamma)$ for strong strain stiffening from Eq.~\ref{eqn:chainprediction} based on stretching of links.
}
\label{fig:stressstrain}
\end{figure}
 
To illustrate the transition from convex particle behavior in the limit of unlinked beads ($N=1$) to strain stiffening in the limit of large $N$, we show stress-strain curves for packings of chains with different lengths $N$ in Fig.~\ref{fig:stressstrain}.   The data fall into two distinct bands:  the upper band of data with chain lengths $N \ge 9$ exhibits strong strain stiffening, while the lower band of data with $N \le 8$  is much weaker.  Packings with chain lengths $5 \le N \le 8$ exhibit mild strain stiffening, while those with $N \le 3$ exhibit the strain-softening typical of convex particles.  While the stress $\sigma$ tends to get larger with increasing $N$, the curves in Fig.~\ref{fig:stressstrain} are not monotonically increasing with $N$, indicating a large run-to-run variation ($\sim 30\%$) compared to any trends in $N$ for $N \ge 9$.

\subsection{Entangled area}
\label{sec:area}

\begin{figure}
\includegraphics[width=3.4in]{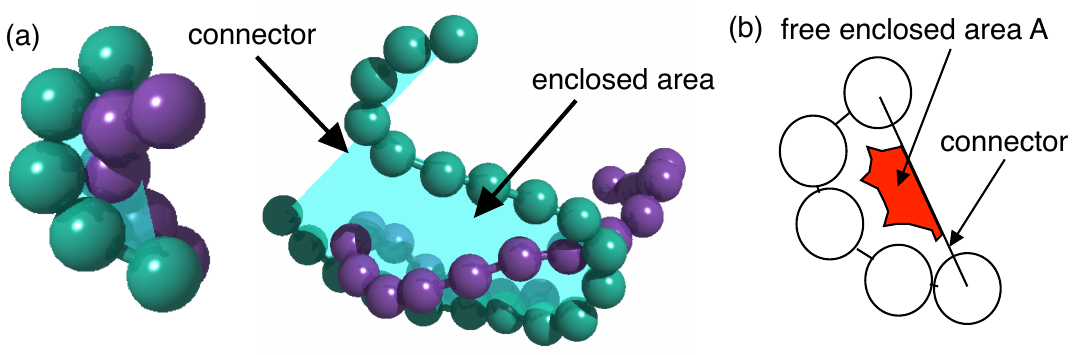}
\caption{(a) Examples of entanglements for chain lengths $N = 5$ (left) and $N =20$ (right) from reconstruction of x-ray tomography images. The green chains are entangling the purple chains. The shaded region indicates the entanglement manifold bounded by the entangling chain and a line connecting the ends of the chains. Examples are reproduced from \cite{BNAJ12}. (b) Illustration of  the free enclosed area $A$ where the center of another chain could pass through the area enclosed by the chain and connector.
}
\label{fig:freeareadef}
\end{figure}

 Since the strong strain-stiffening was found only for chain lengths which have system-filling structures of entanglements  \cite{BNAJ12}, in this subsection we start developing a model to predict the number of entanglements. %by relating the probability of entanglement to the area enclosed by chains in packings 
 We previously presented an algorithm to identify such entanglements in x-ray tomography data  \cite{BNAJ12}.  Motivated by string knotting analysis \cite{RS07}, we first form an enclosure by drawing an imaginary straight  line called the connector between any two beads of the entangling chain separated by a contour length $m$ beads along the contour of the chain.  We then define an entanglement manifold as the minimal two-dimensional area bounded by this enclosure made by the connector and the contour of the chain.  If another chain crosses through this manifold, it is counted as entangled by the entangling chain between the two beads separated by contour length $m$.  Examples of entanglements along with the manifolds are shown in Fig.~\ref{fig:freeareadef}a.  Note that these particular examples would not be counted as entanglements based on typical polymer chain algorithms where entanglements are counted if chains catch on each other if they are contracted \cite{LKMFK09}.  In contrast with polymers chains, our granular chains are highly constrained from significant rearrangements due to their high packing density, so we expect this criteria for entanglement is still able to produce significant constraints against shearing.  %We previously showed that only packings of longer chains which result in roughly at least 0.5 entanglement per enclosure result in clusters of at least half of the chains entangled together, and strong-strain stiffening \cite{BNAJ12}.

%We count all of the entanglements for each possible enclosure between beads and plot the average number of entanglements per enclosure as a function of the contour length $m$ in Fig.~\ref{fig:freeareadef}b \cite{connectornote}.  Regardless of chain length $N$, for contour lengths $m \le 3$, no entanglements are observed because the size of the beads restricts other chains from fitting inside even the largest possible entanglement manifolds.  For larger $m$, the average number of entanglements per enclosure increases monotonically with $m$.   Since chains that are longer compared to their minimum loop size can be bent further, the area of the entanglement manifold tends to similarly increase with $m$ and it is more likely in a random packing that other chains will cross this manifold to become entangled.  Thus, the entanglement manifold method provides an intuitive way to understand why the probability of entanglements increases with separation or chain length.

\begin{figure}
\centerline{\includegraphics[width=3.in]{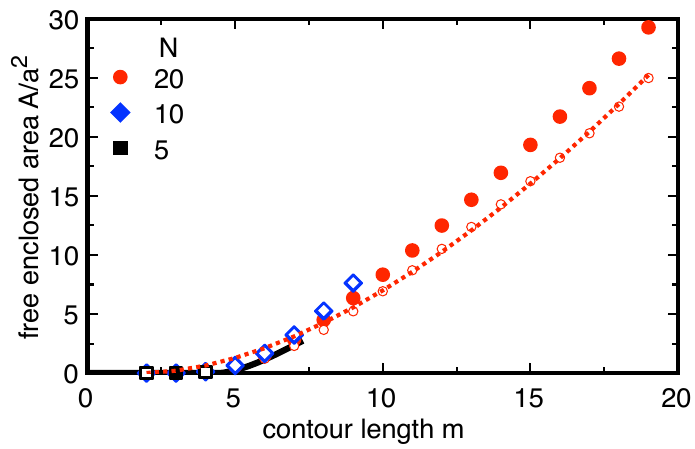}}
\caption{Average free area $A$ enclosed by chains (in units of bead diameter $a$ squared) as a function of contour length $m$. Chain lengths $N$ are given in the legend.   Open symbols: strain $\gamma=0$.  Filled symbols: $\gamma=0.2$. Solid line:  prediction for a semi-circular contour of length $m$ with characteristic circumference  based on the mean bend angle $\langle\theta\rangle$.  Dotted line:  prediction from a random walk model.
}
\label{fig:freeprojarea}
\end{figure}

To develop a quantitative model to predict the probability of entanglements, we first measure the free enclosed area $A$ enclosed by each chain which is available for other chains to pass through.  This is calculated from the x-ray tomography data as the area in the entanglement manifold that is at least 1 bead diameter away from a bead in the entangling chain, as illustrated in Fig.~\ref{fig:freeareadef}b.  This condition accounts for the fact that the finite bead size excludes the center of other beads from overlapping with the beads in the entangling chain, a.k.a.~an excluded volume effect.  The average values for each contour length $m$ are shown in Fig.~\ref{fig:freeprojarea}, shown in units of bead diameter squared.  The shorter contour lengths have very little free enclosed area $A$ because the chains are short compared to the persistence length or minimum loop size and tend to be straighter.  For contour lengths of $m\le 3$, the finite bead size completely excludes chains from passing through the enclosed area for this minimum loop size. This explains why no entanglements were observed for $m\le 3$ \cite{BNAJ12}.   Since chains that are longer can be bent into larger arcs, the free enclosed area tends to increase with $m$, similar to the increase in number of entanglements with $m$ \cite{BNAJ12}.  

%and it is more likely in a random packing that other chains will cross this manifold to become entangled. 

\begin{figure}
\centerline{\includegraphics[width=2.75in]{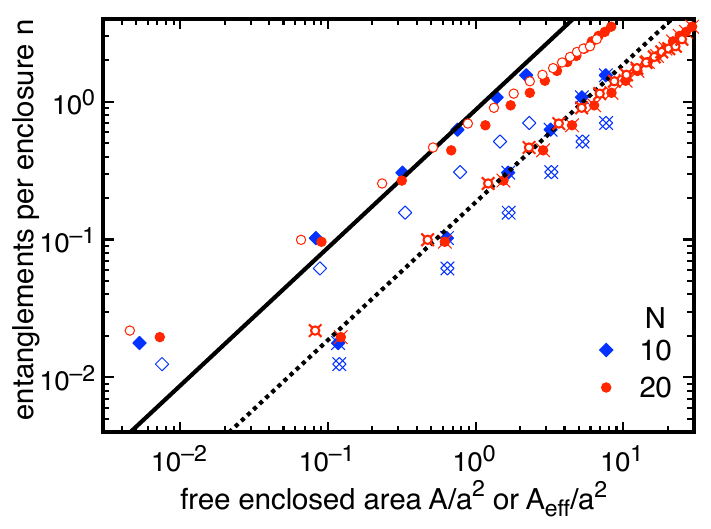}}
\caption{Number of entanglements $n$ per enclosure as a function of free enclosed area $A$.  The points correspond to different contour lengths $m$, using the same data as Fig.~\ref{fig:freeprojarea}.  Circles: chain length $N=20$. Diamonds: $N=10$.  Open symbols: $\gamma=0$.  Filled symbols: $\gamma=0.2$.  Crossed symbols: plotted against free enclosed area $A$. Non-crossed symbols: plotted against effective free enclosed area $A_{eff}$ including constraints in the third dimension.  Lines: linear fits to $N=10$ data.   The near proportionality between $n$ and $A$ or $A_{eff}$ suggests that the probability of entanglement is determined by random arrangements with excluded volume effects, so the average number of entanglements and whether or not there are likely to be system-filling clusters of entanglements can be predicted from particle shapes in the packing.
%The dotted line indicates the threshold for strong strain stiffening based on $N=10$.
}
\label{fig:entanglefreeprojareawangle}
\end{figure}
%self-entanglements?
We plot the average number of entanglements $n$ per enclosure (allowing for multiple entanglements to be counted between each pair and self-entanglements) as a function of free enclosed area $A$ in Fig.~\ref{fig:entanglefreeprojareawangle}.  The different points correspond to measurements for different contour lengths $m$, with larger areas $A$ generally corresponding to longer contour lengths $m$.  A nearly linear relationship between $n$ and $A$ is found (dotted line).  %This proportionality is non-trivial because of the kink in Fig.~\ref{fig:freeprojarea}; the proportionality required that both the probability of entanglements and the free enclosed area start to grow at the same contour length, so that the kinks in both trends  cancel each other out.  
This proportionality suggests that the probability of entanglement is proportional to the free enclosed area $A$, which would be the case for example if the probability of entanglement can be calculated assuming random arrangements limited only by excluded volume. 

%We fit the ratio $n/A$ for each data set in Fig.~\ref{fig:entanglefreeprojarea} yielding values in the range $n/A=0.228 \pm 0.001$ for $N=10$, and $n/A=0.173\pm0.004$ for $N=20$ in the same range of $m$, closer to the threshold for system-filling clusters and strong strain stiffening.  The number of entanglements grows more slowly with larger $A$ (corresponding to larger $m$) presumably due to stronger correlations in particle positions such as excluded volume effects that affect the likelihood of multiple chains crossing through the same entanglement manifold when the average number of entanglements in a manifold is greater than 1.  The linearity for small $m$ suggests $n/A$ can be estimated as an effective two-dimensional packing fraction for chains crossing through entanglement manifolds, with random particle positions so that the probability of a chain crossing through an entanglement manifold is equal to the fraction of the area of entanglement manifolds.  

The number of entanglements $n$ is predicted based on random arrangements with excluded volume effects using a model calculation  that was originally developed for staple packings \cite{GFHG12}.  It involves equating two calculations of volume of entangled particles inside the volume made by an entanglement manifold of thickness $\delta$ -- in this case we set $\delta$ equal to one bead diameter $a$.  The average volume of chains in an entangled volume $V_{ent}$ can be estimated as the average number of entanglements $n$ times the volume per unit length of chain taken up by an entangled bead $\pi a^3/6\bar d$  (with average center-to-center spacing $\bar d$) times the length $\delta/\cos\Theta$ in the entangled plane of thickness $\delta$, where $\Theta$ is the angle of the entangled chain relative to a line perpendicular to the manifold (see illustration in Fig.~\ref{fig:Aeffcalc} of Appendix \ref{sec:Aeffcalc}).  This results in $V_{ent} =  n\pi a^3\delta/6\bar d\cos\Theta$. This is balanced with a prediction of $V_{ent}$ based on the assumption that particle positions are random so the volume of entangled particles is equal to the packing fraction times the free volume available inside the entanglement.  This is calculated as $V_{ent} = \phi A(\Theta,\Phi)\delta$, where $A(\Theta,\Phi)$ is a modification of $A$ which also accounts for the three-dimensional excluded volume from different possible angles $\Theta$ and $\Phi$ around this axis.   Equating the two calculations of $V_{ent}$ results in $n = 6\bar d\phi \langle A(\Theta,\Phi) \cos\Theta\rangle /\pi a^3 = 0.93 \langle A(\Theta,\Phi)\cos\Theta \rangle/a^2$ for $\phi = 0.44 \pm 0.03$  (see Appendix \ref{sec:phi} for measurements of $\phi$) and average center-to-center distance $\bar d/a  = 1.11$ (measured from x-ray tomography).  The effective free enclosed area $A_{eff} = \langle A(\Theta,\Phi) \cos\Theta\rangle$ accounts for excluded volume effects in three dimensions, as well as the larger cross sections of entangled chains at various angles $\Theta$.   $A_{eff}/A$ corresponds to the parameter $1/\alpha$ introduced as a fit parameter by \cite{GFHG12}.  The calculation of $A_{eff}$ from our x-ray tomography data is explained in Appendix \ref{sec:Aeffcalc}.  The number of entanglements $n$ as a function of $A_{eff}$ calculated from x-ray tomography data is shown in Fig.~\ref{fig:entanglefreeprojareawangle}.  This also appears nearly linear, as there is nearly a constant ratio $A_{eff}/A= 0.27\pm0.01$ in the strong strain-stiffening range with $m\ge 8$ (see Appendix \ref{sec:Aeffcalc}).  For the $N=10$ data in Fig.~\ref{fig:entanglefreeprojareawangle} with percentage errors,  a linear fit yields $n = (0.87\pm0.14) A_{eff}/a^2$ and a coefficient of determination of 0.92, confirming a nearly linear trend as predicted, with a coefficient consistent with the predicted value of $n=0.93A_{eff}/a^2$. 
%A linear fit yields $n = (0.74\pm0.02) A_{eff}/a^2$ for $N=10$ from Fig.~\ref{fig:entanglefreeprojareawangle} with constant errors, with a coefficient of determination of 0.99, confirming that a linear relation is a good approximation of the data,  The fit coefficient is close to the predicted value of $n=(0.93 \pm 0.06)A_{eff}/a^2$.  
This confirms that the assumption of random positions and orientations with excluded volume effects can relate the number of entanglements $n$ to the free enclosed area $A_{eff}$ if the shape of the particles in the packing is known, at least for $N=10$.   The measured  ratio $n/A_{eff}$ decreases for $N=20$ and for increasing $m$, suggesting that longer chains have additional constraints  that reduce the number of entanglements, for example due to the maximum bend angle of chains which puts constraints on particle positions in the same chain.  Since $n/A$ is more linear than $n/A_{eff}$ for small $m$, the larger $n/A_{eff}$ for small $m$ with the assumption of random particle orientations $\Theta$ in $A_{eff}$ indicates the angles $\Theta$ are more perpendicular to the entangled plane than random orientations, which could be a result of contact forces tending to push entangled chains into the small range of angles that they will fit.
%For the staples of \cite{GFHG12}, $w=1.17$ cm, thickness $D=0.5$ mm, 

%10mers_compressued missing 9% of beads 
%20mers_compressed missing 16.4% of beads - may cause underestimate of $n/A$ by up to this amount

\subsection{Prediction of system-filling clusters from free enclosed area}
\label{sec:clusters}

A prediction of the free enclosed area $A_{eff}$ would extend the prediction of $n/A_{eff}$ to predict the number of entanglements $n$ and whether or not a particular particle shape is likely to have system-filling clusters of entanglements.  To predict the effective free enclosed area $A_{eff}$ it is more intuitive to work with a calculation of $A$ in two dimensions, and treat $A_{eff}/A= 0.28\pm0.01$ as a constant parameter based on the known particle geometry in the packing (see Appendix \ref{sec:Aeffcalc}) to convert to $A_{eff}$.  

To predict $A$ for flexible chains, we calculate the area enclosed by a characteristic chain shape.  We choose as one limiting case a semicircular arc since there is a tendency for chains to form loops.  In a two-dimensional plane, a semicircular arc of contour length $m$ has circumference of $c= 2\pi m/\langle\theta\rangle$ where  $\langle\theta\rangle= 25^{\circ}$ is the mean bend angle between adjacent links in the chain, obtained from x-ray tomography data.   The free enclosed area for such a semicircular arc  with $m < c/2$ can be expressed as $A_s= (\theta_f/2) (r-1)^2 - r(r-1) \cos(\langle\theta\rangle/2)\sin(\theta_f/2)$ where the radius $r = c/2\pi$ and $\theta_f/2 = \arccos[r\cos(\langle\theta\rangle/2)/(r-1)]$ for $m<c/2$.   This predicted area $A_s$ is shown as a function of contour length for $m\le 7$ in  Fig.~\ref{fig:freeprojarea},  since the autocorrelation of bend angles drops to zero by $\Delta m=7$  (see Fig.~\ref{fig:bendanglecorr} in Appendix \ref{sec:chainbending}).  %The minimum contour length to enclose free area greater than zero is predicted to be $m_{min} = (c/\pi)\arccos(1-2\pi/c)$.  
There is good agreement with the minimum contour length for non-zero free enclosed area, and the predicted free enclosed area $A_s$ is within the scatter of different measurements up to $m=7$.   This agreement indicates that the free enclosed area and thus the probability of entanglement can be predicted near the critical threshold $m_{min}$ based on a characteristic chain shape with a single geometric parameter such as mean bend angle $\langle\theta\rangle$.  %Since the enclosed area is not highly sensitive to the mean bend angle, this behavior could even be estimated from a parameter such as the minimum loop size without a measurement of the mean bend angle.
%Since this calculation was done in a two-dimensional plane calculating the excluded area as if the entangled bead is centered in the same plane, it could in principle overestimate the excluded area, but the agreement with data suggests that the overestimate is not significant. 

For $m>7$ where the bond angles are no longer correlated, the assumption of a highly-correlated semicircular arc is no longer very sensible.  When the contour length is  large compared to the bend angle correlation length, an uncorrelated random walk might be a more appropriate. %A power law with an offset $m_{min}$ is fit to data for $N=20$, $\gamma=0.2$ which has an exponent $1.32\pm0.02$, not quite at the level of a wormlike chain.
Assuming a random walk results in an average distance $(\bar d m'/a)^{0.5}$ away from the connecting line up to $m/2$ and mirror imaged to return to the connecting line results in a free enclosed area $A_{rw}/a^2=(2\bar d/a)\int_{a/\bar d}^{m/2}[(\bar d m'/a)^{0.5}-1]dm' =\sqrt{2}(\bar d m/a)^{1.5}/3-\bar d m/a+2/3$.  The $-1$ term and starting the integral at $m'=a/\bar d$ where the integrand equals 0 accounts for the excluded volume effect.  This random walk prediction $A_{rw}$ is shown in Fig.~\ref{fig:freeprojarea}, which is within the scatter of the data for the  chains long enough to entangle ($m\ge 8$). %The free enclosed area $A$ from the random walk model differs from the semicircular arc by only an average of $1.0a^2$ for $m \le 7$, which has only a small effect on the subsequent prediction of system-filling clusters of entanglements.
For small $m$, a random walk predicts an unrealistically high value for $A$ because it ignores the limitation on bend angles.  Thus, a random walk model predicts the free enclosed area $A$  for chain longer than the bend angle correlation length,  while a semicircular arc model predicts the number of entanglements for shorter chains with more correlated bend angles.

%erdos renyi
The observed transition to strong strain stiffening at $N=9$ corresponding to the rapid growth of system-filling clusters around $m=8$ \cite{BNAJ12} can be understood by considering that the free enclosed area $A/a^2=4.4\pm0.8$ at $m=8$  in Fig.~\ref{fig:freeprojarea} corresponds to a threshold number of entanglements per enclosure of $n=0.8\pm0.2$ in Fig.~\ref{fig:entanglefreeprojareawangle}.    This threshold number of entanglements is similar to the Erd{\"o}s-R{\'e}nyi random graph model, which is also used to model gelation of polymer networks based on the fraction of bonds or entanglements.  In the analogy to the Erd{\"o}s-R{\'e}nyi model, each chain is represented by a node in a graph with an equal probability of being connected (i.e.~entangled) to each other node.   There is a sharp increase in the fractional size of the largest cluster around an average number of connections per node of 1, corresponding to entanglements per enclosure around 1 \cite{ER59}.  

%generalization to staples
This threshold of $n=1$ entanglements per enclosure is consistent with some previous experiments with staples. The average number of entanglements per enclosure $n$ was calculated for staples as a function of aspect ratio \cite{GFHG12}.  The calculation crosses the expected threshold of 1 for system-filling clusters of entanglements from the Erd{\"o}s-R{\'e}nyi model at an aspect ratio of arm length to backbone length of $0.1$ \cite{GFHG12}.  This matched with the smallest aspect ratio that was experimentally found to consistently support freestanding structures, while experiments at a smaller aspect ratio of 0.02 often resulted in a partial collapse when the confining boundary was removed \cite{GFHG12}.  This ability to support stresses larger than the scale of the confining stress at the boundary to support freestanding structures is one of the features of strong strain stiffening, as it implies the packing strength exceeds the confining stress.  These observations are consistent with a general argument that an average greater than about $n=1$ entanglements per enclosure leads to system-filling clusters of entangled particles, which can provide enough constraints from entanglement to prevent the usual failure by shear banding.  If we take $n=1$ entanglement per enclosure as an approximate general threshold for system-filling clusters and strong strain stiffening, then whether this threshold is crossed can be predicted for different particles shapes by calculating the free enclosed area $A$ and $A_{eff}$ to get $n$ using the  techniques in Sec.~\ref{sec:area} and Appendix \ref{sec:Aeffcalc}, assuming the particle shape in the packing is known.

\subsection{Stretching of links}
\label{sec:linklength}

\begin{figure}
\centerline{\includegraphics[width=3.5in]{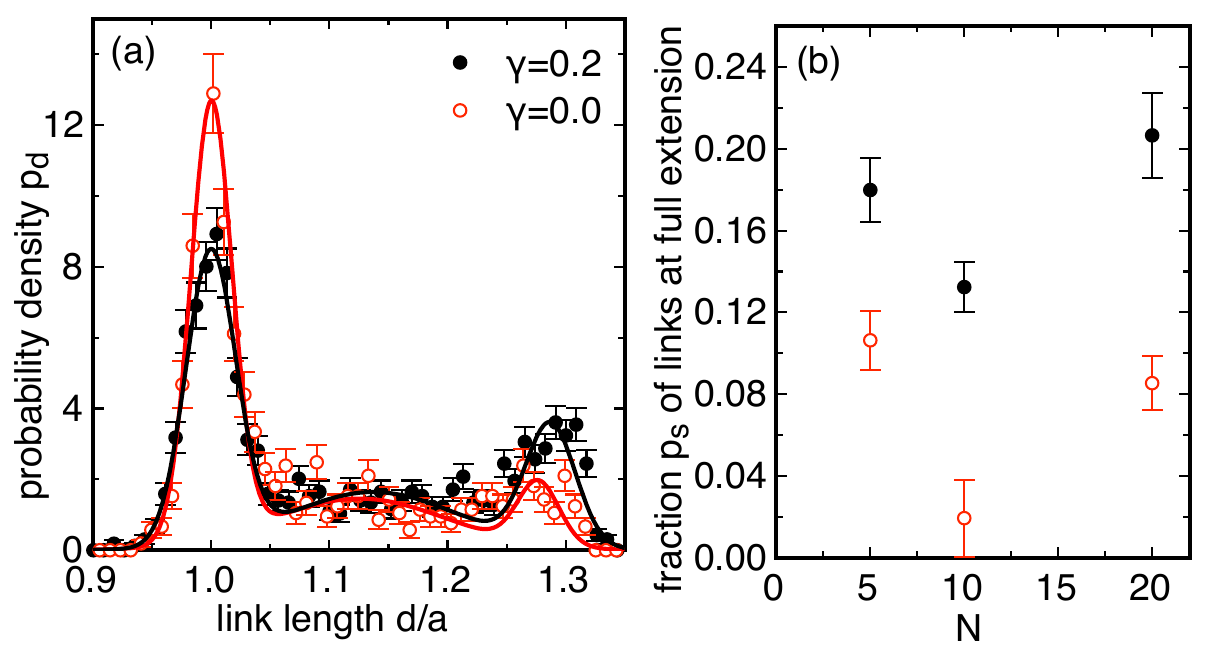}}
\caption{(a) Distribution of center-to-center distances $d$ among neighboring beads in a chain normalized by bead diameter $a$ for $N=20$ for strains $\gamma = 0.0$ (open symbols) and $\gamma=0.2$ (Filled symbols).  Solid lines:  fits of 3 Gaussian peaks to the data.  (b) Fraction of fully-extended links in rightmost peak of panel a based on Gaussian integrals.  The increase in the number of fully extended links indicates that the chains are being stretched when the packing is strained, and the shift in the peak position indicates the average strain on those links, which results in a significant load on the stretched links.
%Circles: Fraction $p_s$ of stretched links in rightmost peak at $d/a>1.25$ and $\mu>0.80$, corresponding to the nearest neighbor bead contacting each bead on the end of the link.  Compression of the packing is seen to fully stretch more links.
}
\label{fig:linkdist}
\end{figure}

To determine whether individual chains become stretched or compressed in strained packings, we measure the distance $d$ between neighboring beads in each chain from x-ray tomography.  The normalized probability distribution of center-to-center distances $p_d(d/a)$ is shown in Fig.~\ref{fig:linkdist}a for $N=20$ for global strains $\gamma = 0$ and $0.2$.   The large peak on the left of Fig.~\ref{fig:linkdist}a is at the bead diameter $a$, where many of the beads are pressed against each other in contact.  The smaller peak on the right of Fig.~\ref{fig:linkdist}a corresponds to links at maximum extension before particle deformation occurs.   This peak is at a length slightly  less than the center-to-center distance for straight links $d_0$, which occurs because the geometry of the links connecting beads is such that the maximum center-to-center distance decreases with chain bend angle $\theta$.
%(see Fig.~\ref{fig:bendangledist}) 
%The average width of this peak is measured to be $0.044 \pm 0.004$ mm (the uncertainty is the standard deviation of the mean over the six data sets), in agreement with the standard deviation of the average center-to-center distance at maximum extension  $\sigma_{d_0} = 0.045\pm 0.005$ mm.  This suggests that the peak width is due to the natural variation of the bead diameters, and all of the points in the large peak of Fig.~\ref{fig:linkdist} correspond to beads in contact.  

The relative sizes of the two peaks in Fig.~\ref{fig:linkdist}a indicates how many links are compressed or fully extended.  To distinguish the peaks from the background, the distribution $p_d$ is fit to the sum of three Gaussian functions -- the left peak for compressed chains, the right peak for fully extended chains, and a broad middle Gaussian for the background.  The fraction of fully extended links $p_s$ is calculated as the integral of the Gaussian fit to the rightmost peak of the probability distribution $p_d(d/a)$ and shown in Fig.~\ref{fig:linkdist}b for each $N$ and $\gamma$.    
%The fraction of links at full extension increases by $0.119 \pm 0.025$ as the packing is strained from $\gamma=0$ to $\gamma=0.2$. The uncertainty of 0.025 is the same for both the uncertainty propagated from the fit parameters, and the standard deviation of the three samples, indicating that the number of links stretched in each sample is consistent within the uncertainty.  
This indicates a significantly increased fraction of stretched links when the packing is strained, and the change in the fraction of extended links is similar for different $N$.  

%Further analysis suggests that these stretched links in strained packings are more likely to have a bead from a neighboring chain pressed up against the two beads of the link (see Fig.~\ref{fig:mudst}).
%measurements of particle strain
%The particle strain $\gamma_p$ can be estimated from Fig.~\ref{fig:linkdist}.  Since the additionally stretched links under tension are the ones that can help hold the load, we assume for simplicity those center-to-center distances go all the way from $a$ to $d_0=1.29a$, while no others are strained since the intermediate probability distribution remains similar,  and the fraction of locally strained particles increases by 0.119 from $\gamma=0$ to $\gamma=0.2$,  that corresponds to a contribution to particle strain $\gamma_p/\gamma =0.18$.  Similarly, the average change in center-to-center distance from $\gamma=0$ to $\gamma=0.2$ for the samples that started off dense $N=5, 20$, corresponding to a $\gamma_p/\gamma=0.17$.   

%particle strain measurements
 While the integral of the rightmost peak of Fig.~\ref{fig:linkdist}a indicates the fraction of links that are fully extended so that they may stretch, the average strain $\gamma_p$ of those links that are stretched can be determined from the change in position of the rightmost peak.  Comparing the positions $d_R$ of the rightmost peaks in Fig.~\ref{fig:linkdist}a indicates an additional link strain $\gamma_p = [d_R(\gamma=0.2)-d_R(\gamma=0)]/d_R(\gamma=0)= 0.016\pm 0.002$ for $N=20$.  This average strain corresponds to an average tensile load of $44\pm6$ N on each stretched link (see Fig.~\ref{fig:chainextension} in Appendix \ref{sec:chainstiffness}).  This is a significant fraction of the average total load of 1400 N on strong strain stiffening samples at $\gamma=0.2$, which suggests that stretched links could support a significant part of the load in strong strain stiffening samples \footnote{We could not calculate the strain for $N=10$ because there was not a clearly resolvable peak in $p_d$ at $\gamma=0$, most likely due to that packing being initially very loose (see Fig.~\ref{fig:phi} in Appendix \ref{sec:phi}).}  In contrast, for $N=5$, we measure a $\gamma_p =  0.001\pm0.002$, consistent with zero and much smaller than the value for $N=20$.  This confirms that there is no resolvable stretching of links for the chains that are too short to have system-filling networks of entanglements at $N=5$.  

%compression of loops
Considering bending of loops, we find that the average bend angle of chains decreases with increasing strain $\gamma$, and the largest compressive force on a tight loop is only 4 N (see Appendix \ref{sec:chainbending}), much less than the average load on stretched links, suggesting that the stress from compression of loops is insignificant compared to the stress from stretching of links in strong strain stiffening of chains \cite{LRR11}.

%\section{Autocorrelation of center-to-center distances}
%\label{sec:correlationlength}

\begin{figure}
\includegraphics[width=2.8in]{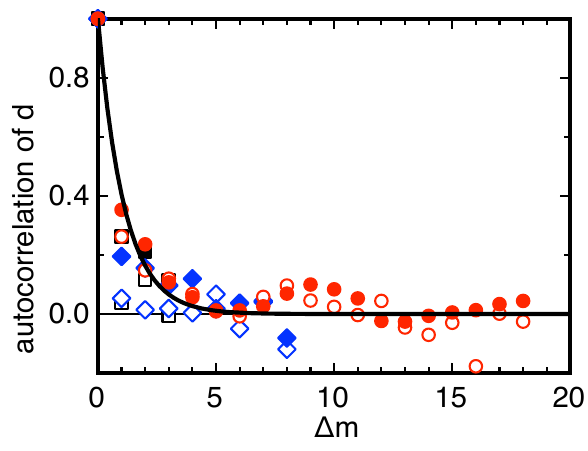}
\caption{The autocorrelation of center-to-center distance $d$ along a chain separated by $\Delta m$ beads.  Circles: $N=20$.  Diamonds: $N=10$.  Squares: $N=5$.   Open symbols $\gamma=0$.  Filled symbols: $\gamma=0.2$.  The small correlation length suggests force chains are not primarily transmitted along the length of chains. 
}
\label{fig:linkdistcorr}
\end{figure}

%Figure \ref{fig:linkdistcorr} shows the autocorrelation of center-to-center distances $d$  as a function of the separation distance $\Delta m$ along the chain in units of beads.  There is a significant variation of these correlation lengths among the packings at $\gamma=0$, suggesting the initial packing structure might affect this.  An exponential fit to both $N=10$ and $N=20$ data at $\gamma=0.2$ for $m\le 5$ yields a characteristic correlation length of $\lambda=1.1\pm0.1$ beads. %This short correlation length suggests that only nearest neighbor beads in a chain are dominant in understanding the geometry of entanglements in terms of nearby bead positions.
%$\lambda=1.11\pm0.03$ for $N=20$. 
%$\lambda=1.00\pm0.13$ for $N=10$.

%interpretation of lambda=1
%The unresolvable orientation of individual dog-bone shape of links and that orientation's effect on link strain introduces too much variation from link to link to identify precise forces on individual particles, but statistics are still telling.  
The autocorrelation of center-to-center distances $d$ is shown as a function of the separation distance $\Delta m$ along a chain in units of beads in Fig.~\ref{fig:linkdistcorr}.  There is a significant variation of correlation lengths among the packings at $\gamma=0$, suggesting the initial packing structure might affect this.  An exponential fit to both $N=10$ and $N=20$ data at $\gamma=0.2$ for $m\le 5$ yields a characteristic correlation length of $\lambda=1.1\pm0.1$ beads.  This short correlation length suggests that entangled chains do not consist of a sequence where every link in the same chain is fully extended, i.e.~with the force chain following the links of the chain.  Rather, the stretched links can still act as extra constraints on the packing  where large forces can be supported, but these forces may be transmitted along frictional contacts with other beads instead of just along the links of a single chain. 
%coordintaion number: jamming with friction 4

\subsection{Prediction of the stress-strain curve}
\label{sec:model}

Now that we have established that strain stiffening occurs along with stretching of links, but only in packings with system-filling clusters of entanglements, we desire a predictive model of how that leads to strong strain stiffening.  Since the stiffness of the beads is much higher than that of the elastic boundary, the stiffness of the overall system would be limited by the weaker  stiffness of the boundary without constraints that can provide support across the entire system.  Importantly, system-filling clusters of entanglements allow for a possible way to provide additional internal constraints that prevent failure by shear banding and provide an effective replacement for confinement at the boundary.  With such constraints, the strength can exceed the confining pressure at the boundary by stretching of the chains.   
While the strength can originate from the stretching of chains, the constraints of entanglement may allow forces to be transmitted along a contact network that includes the entire system when connected by system-filling clusters of entanglements.
%We assume that magnitudes of stresses on links will transmit from particle to particle along a force network that redirects forces such that the average magnitude of stress on the endplates is similar to the average magnitude of stresses on the particles.

To calculate the stress-strain relation  we use a mean-field mode where we assume the average stress $\sigma$ measured on the endplate in response to triaxial load ultimately comes from and is equal to a volume average of the  magnitude of the stress on a cross-section of particles.  This matching of magnitudes regardless of the direction of forces is reasonable if those contact forces are redirected randomly along a force network.   To relate the global stress $\sigma$ to the particle stress $\sigma_p$ and particle strain $\gamma_p$, we write the derivative of the global stress-strain relation using the chain rule from calculus
\be
\frac{\partial \sigma}{\partial \gamma} = \left<\frac{\partial \sigma}{\partial \sigma_p}\right>\left<\frac{\partial \sigma_p}{\partial \gamma_p}\right> \left<\frac{\partial \gamma_p}{\partial \gamma}\right> \ .
\label{eqn:dtaudgamma}
\ee
The angle brackets represent a mean-field average to be calculated as a volume average over a typical local structure.   The effective extensional modulus of the chain $\partial \sigma_p/\partial \gamma_p \equiv E_p$ is calculated as the force pulling on the chain divided by the average cross-sectional area of the chain and the strain of the chain (see Appendix \ref{sec:chainstiffness}).  $\langle\partial\sigma/\partial\sigma_p\rangle$ is the ratio of average stress over the entire volume to the average stress on a stretched chain, corresponding to the fraction of the volume that is being stretched, and is calculated as the packing fraction $\phi$ times the probability of stretched links $p_s$.     $\langle\partial \gamma_p/\partial \gamma\rangle$ then relates the particle strain to the applied strain $\gamma$ averaged over links that are stretched.   Equation \ref{eqn:dtaudgamma} can now be rewritten as 

\be
\frac{\partial \sigma}{\partial \gamma} =  \phi p_s(\gamma) E_p \left<\frac{\partial \gamma_p}{\partial \gamma}\right> \ .
\label{eqn:dtaudgammastretch}
\ee
In this formulation, $E_p$ is a constant for elastic particles at small strain $\gamma$ (see Appendix \ref{sec:chainstiffness}), and $\phi$ is roughly constant to within a few percent (see Appendix \ref{sec:phi}).     The particle strain ratio $\langle\partial \gamma_p/\partial \gamma\rangle$ term is dependent on local structure, but for small packing deformations, we can assume the local structure does not evolve significantly with strain, so this term is should be nearly constant, at least in the limit of small strain.  %The details of this calculation are shown in Sec.~\ref{sec:localstrain}.   
Only the probability of chains being stretched $p_s(\gamma)$ is expected to increase significantly with strain $\gamma$, as seen in Fig.~\ref{fig:linkdist};  as the system is further strained, more and more entanglements should be pulled tight and stretched.  Thus we hypothesize this $p_s(\gamma)$ term in Eq.~\ref{eqn:dtaudgammastretch} to be responsible for the increase in $\partial\sigma/\partial\gamma$ and  strain stiffening.

%p_s
We estimate $p_s(\gamma)$ and $\langle\partial \gamma_p/\partial \gamma\rangle$ based on the x-ray tomography measurements of center-to-center distance $d$, shown in Fig.~\ref{fig:linkdist}.      Under strain, the change in the fraction of links at full extension is $ 0.12\pm0.02$ at $\gamma=0.2$ for both $N=10$ and $N=20$ (Fig.~\ref{fig:linkdist}b).  Assuming to lowest order that $p_s$ is proportional to strain, this corresponds to
\be
p_s(\gamma) =  (0.6\pm0.1)\gamma  \ .
\label{eqn:probstretchval}
\ee

The measured average particle strain is $\gamma_p= 0.016\pm 0.002$ at $\gamma=0.2$ for $N=20$ based on Fig.~\ref{fig:linkdist} (see Sec.~\ref{sec:linklength}).  Assuming $\langle\partial \gamma_p/\partial \gamma\rangle$ is a constant for small strain, this corresponds to $\langle\partial \gamma_p/\partial\gamma\rangle = 0.080\pm0.012$.

%combined model
Plugging into Eq.~\ref{eqn:dtaudgammastretch} the measured value of the extensional modulus $E_p = (1500\pm90)$ MPa from the fit of  Fig.~\ref{fig:chainextension} in Appendix \ref{sec:chainstiffness}, the measured $\phi  \approx 0.43\pm0.02$ for the $N=10$ and 20 packings from Fig.~\ref{fig:phi},  $p_s(\gamma) =  (0.6\pm0.1)\gamma$ from Eq.~\ref{eqn:probstretchval}, and the measured $\langle\partial \gamma_p/\partial\gamma\rangle = 0.080\pm0.012$ based on Fig.~\ref{fig:linkdist}, 
results in the prediction $\partial\sigma/\partial\gamma=(31\pm7)\gamma$ MPa.  Integrating yields 
\begin{equation}
\sigma = (16\pm4)\gamma^2 \mbox{ MPa} \ .
\label{eqn:chainprediction}
\end{equation} 
%$\partial\sigma/\partial\gamma=(31.0\pm7.2)\gamma$ MPa, 

\begin{figure}
\centerline{\includegraphics[width=2.8in]{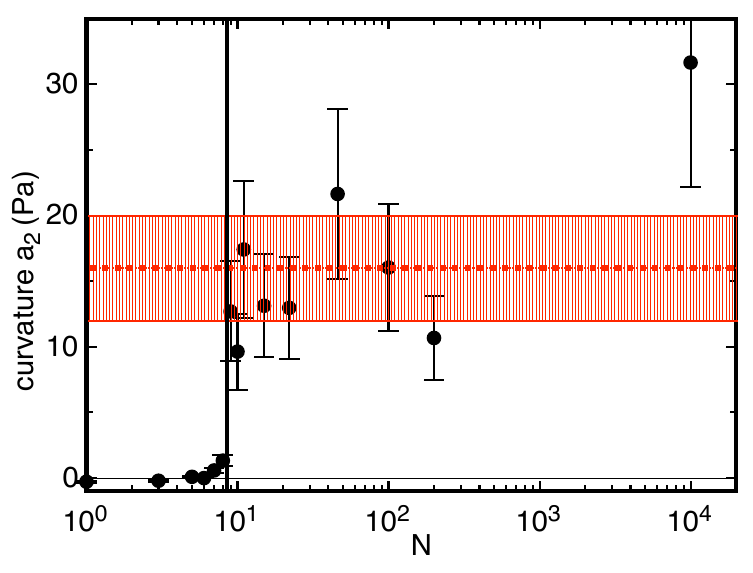}}
\caption{(color online) Fitted curvature $a_2$ of the stress-strain data from Fig.~\ref{fig:stressstrain}, including the 30\% run-to-run variation in the uncertainty.  Dotted line: model prediction for $a_2$ due to stretching of links with system-filling clusters of entanglements.  The model prediction is consistent with each curvature for $N\ge 9$ where strong strain stiffening and system-filling clusters of entanglements are observed.  The vertical line delineates the boundary between weak and strong strain stiffening based on the magnitude of the curvature.}
\label{fig:quadfit}
\end{figure}
%with  $a_1 = 69p_0$ and $a_2 = 21\pm5$ MPa.  %A different threshold of $\mu$ by $\pm0.02$ increases the uncertainty on $a_2$ by 5 MPa.  
%with $a_1 = 62p_0$ MPa and $a_2 = (18.7\pm 3.8)$ MPa based on predicted structure factor
For comparison to the model, a quadratic function $\sigma = a_2\gamma^2$ was fit to each measured $\sigma(\gamma)$ in Fig.~\ref{fig:stressstrain} up to $\gamma= 0.2$.   Fit values of the curvature $a_2$ are shown in Fig.~\ref{fig:quadfit}.
%mean and standard deviation of $a_2 = 16 \pm 7$ MPa.  
For each $N\ge9$ where strong strain stiffening is found, the predicted curvature $a_2$ based on structural measurements with a 25\% uncertainty is consistent with each fit including the 30\% run-to-run variation on the measurements. %although the weak trend in $N$ for $N\ge 9$ is not captured by the coarse model.  
The average coefficient of determination for the quadratic fits for $N\ge 9$ is 0.988.  Fitting the curve at $N=10000$ up to a larger strain of 0.45 results in less than a 2\% change in $a_2$, confirming the scaling is similar over a wider parameter range.  Since the curvature $a_2$ of the measured $\sigma(\gamma)$ for $N\ge 9$ is consistent with the model prediction, the strength of strain stiffening can be attributed to the stretching of links.   The quadratic shape of strain stiffening curves is attributed to an assumed linear increase in the fraction of links being stretched $p_s$ with strain. 
%2nd order fit with errors from Instron can lead to weighting with poor fit of large strain

%higher order terms due to variation in other parameters
The model of Eq.~\ref{eqn:dtaudgammastretch} has no explicit dependence on chain length $N$, and we measure negligible structural differences from x-ray measurements to indicate that the structural terms would change significantly with chain length $N$, other than a slight weakening due to the decrease in packing fraction $\phi$ with $N$ (Fig.~\ref{fig:phi}).  This is consistent with the observation in Figs.~\ref{fig:stressstrain} and \ref{fig:quadfit} that there is no strong trend in the shape or strength of the stress-strain curve with $N$ for for $N\ge9$ where chains long enough to exhibit strain stiffening.  There is a mild trend of increasing strength with chain length for $N\ge9$ --  due to the $N\approx10^4$ chain --  that suggests $p_s(\gamma)$ or $\langle\partial\gamma_p/\partial\gamma\rangle$ could increase with chain length, but a lack of consistent ordering of strength with chain length for $N \ge 9$ in Fig.~\ref{fig:stressstrain} suggests most of this variation is random, for example it could be due to differences in the initial packing from run to run. % and are not testable with our limited x-ray measurements.  

Given the strain ratio $\langle\partial \gamma_p/\partial\gamma\rangle$, the failure of the packing can also be predicted.   The average link breaking is measured to occur at an average link strain $\langle\gamma_p\rangle=0.05$ (Fig.~\ref{fig:chainextension} in Appendix \ref{sec:chainstiffness}).  Using the measured $\langle\partial \gamma_p/\partial\gamma\rangle = 0.080\pm0.012$ and extrapolating  suggests roughly 50\% of the stretched links would break by packing strain of $\gamma = 0.6$.  Since Eq.~\ref{eqn:probstretchval} suggests 36\% of links are stretched at $\gamma=0.6$, an overall 18\% of links are expected to be broken in the packing at this point.  This is a little more than the 10\% of links that we observed to be broken after a failure observed at $\gamma=0.6$ in Fig.~\ref{fig:longchain} \cite{BNAJ12}, quite close to the predicted failure strain and number of broken links.  Such a prediction should be applied with caution, as it is based on extrapolation of the assumed constant ratio $\langle\partial \gamma_p/\partial\gamma\rangle$ to large strains, and failure of the packing of the whole could depend on weakest link statistics rather than a fixed fraction of broken links.
%The difference in maximum cluster size for packings that cross the threshold to strong strain stiffening was found to be 12\% \cit{BNAJ12}, but only for compressed samples.

\subsection{Scaling behavior}

%linear fiit
\begin{figure}
\centerline{\includegraphics[width=3.2in]{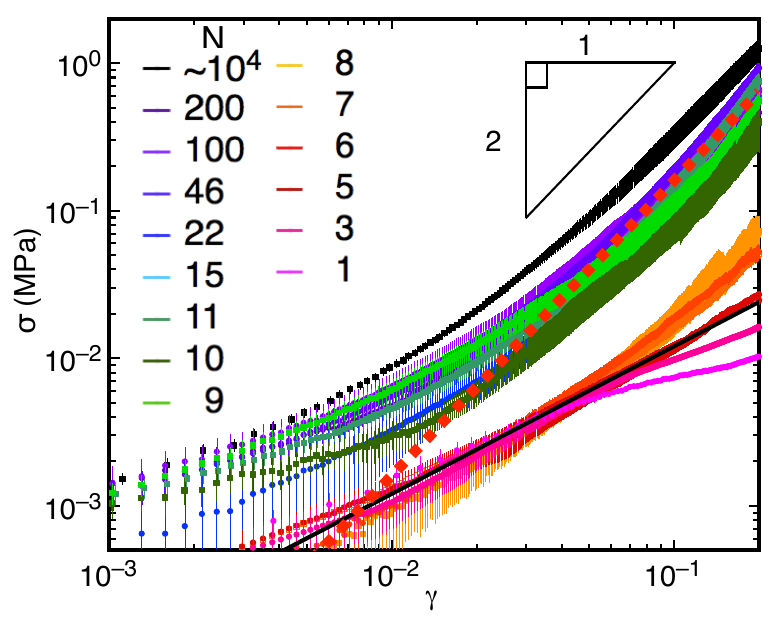}}
\caption{(color online) Stress $\sigma$ vs.~strain $\gamma$ for the same data from Fig.~\ref{fig:stressstrain} on a log-log scale for different chain lengths $N$. Red dotted line: the predicted quadratic stress-strain relation (slope 2) for strong strain stiffening due to system-filling clusters of entanglements. The curves for $N\ge9$ have a slope near 2 for large strain, indicating that they follow the quadratic prediction where there is strong strain stiffening.  Black solid line: slope 1 for reference, highlighting the transition from strain softening to mild strain stiffening at $N\approx 5$.
}
\label{fig:stressstrainloglog}
\end{figure}
To confirm whether the measured stress-strain curves $\sigma(\gamma)$ follow the predicted quadratic curvature, we replot each $\sigma(\gamma)$ from Fig.~\ref{fig:stressstrain} on a log-log scale in Fig.~\ref{fig:stressstrainloglog}.    The strong strain-stiffening curves with $N\ge9$ are seen to be parallel to the predicted quadratic curve with slope 2 for strains $\gamma\stackrel{>}{_\sim}0.04$, showing that the quadratic prediction correctly describes these data where strong strain stiffening occurs.    A transition from strain softening to mild strain stiffening and then strong strain stiffening can be seen for the curves as $N$ increase, which gradually approach a slope near 2 for large $N$.  Thus, the transition from mild to strong strain stiffening is not defined by a sharp change in the scaling law, but rather corresponds to a sharp jump in the magnitude of the curvature of the stress-strain relation between $N=8$ and 9, as shown quantitatively in Fig.~\ref{fig:quadfit}.  

The stress-strain curves in Fig.~\ref{fig:stressstrainloglog} are nearly linear for small strain $\gamma\stackrel{<}{_\sim}0.01$.  This is in a range of stress that is consistently less than 10 kPa, which does not reach significantly higher strength than packings of individual beads ($N=1$, see inset of Fig.~\ref{fig:longchain}).  Since the strength of unlinked bead packings is understood to be determined mainly by the stiffness of the confining elastic membrane \cite{LW69}, much of the linear regime for larger $N$ is also expected to be due to the stiffness of the elastic membrane as well.  Since this linear range of stress is less than 1\% of the stress range that we measure, and our resolution is very limited here, we cannot distinguish the contribution of the elastic membrane from other mechanisms for the linear stress-strain region in the chain packings for $N>1$.  %The orders-of-magnitude increase in the stress 

In Eq.~\ref{eqn:probstretchval}, we could have assumed a constant added to $p_s$ corresponding to the fraction of links fully stretched at zero strain, which could be as large as 0.09 based on the $N=20$ data in Fig.~\ref{fig:linkdist}b.  This would have added a linear term to the stress-strain relation, corresponding to the nearly linear regime in Fig.~\ref{fig:stressstrainloglog} for small strain.  To test if this nearly linear stress-strain regime is related to $p_s$ at zero strain, we added a linear term to the polynomial fit  $\sigma =  a_1\gamma +a_2 \gamma^2 $ to the data. Fits of the stress-strain relation yield an average $a_1 = 0.3\pm0.2$ MPa, which would correspond to a probability of stretched links of $p_s= a_1/ E_p\phi\langle\partial \gamma_p/\partial\gamma\rangle  = 0.004\pm0.003$.  This is much smaller than the number of links that are fully extended at $\gamma=0$ as seen in Fig.~\ref{fig:linkdist}, suggesting that most of these initially extended links are not stretched enough for the link stiffness to contribute to the packing strength.   Furthermore, much of the contribution to $a_1$ is due to the stiffness of the elastic membrane, so the fraction of stretched links at zero strain should be even less than the value inferred from $a_1$, and we should not use the value of $a_1$ to directly infer a fraction of stretched links.  It is reasonable to understand that the fraction of stretched links at zero strain is negligible, as long as the initial packing is done without a significant loading.

\section{Generalizing the strain stiffening model to staples}
\label{sec:staples}

Equation \ref{eqn:dtaudgammastretch} represents a simple model which attributes the large stress in entangled structures to the deformation of particles, with strain stiffening due to an increasing number of deformed particles with strain.  This concept could be extended to a variety of entangled particle systems, not just limited to chains.   To test the generality of this model for strain stiffening we consider a very different particle shape: rigid U-shaped staples, which can also entangle.  Staples have a few features that are helpful for testing the stress-strain model beyond simply having another shape as a data point.  Since the particles are rigid, geometric calculations can be made to determine the free entangled area and predict a  probability of deformation $p_s$ without need for measurements.  Because the particles deform plastically,  we can measure the bend angles of staple arms after compression of packings  to infer information about entanglement and extensional forces between particles without the need for x-ray tomography, and thus obtain data more easily, which is helpful for example to test the strain dependence of factors such as $p_s$ and $\langle\partial\gamma_p/\partial\gamma\rangle$.

\subsection{Strain stiffening}

\begin{figure}
\centerline{\includegraphics[width=2.6in]{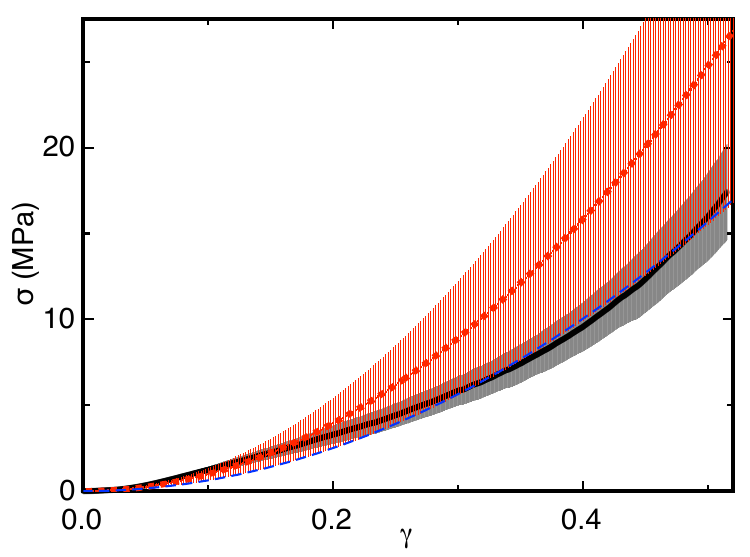}}
\caption{(color online) Another example of strain stiffening in randomly packed particles that can entangle -- in this case rigid U-shaped staples.  The black curve indicates the mean and the gray band indicates the standard deviation of three measurements.  The red dotted line shows the model prediction of Eq.~\ref{eqn:modelstaples} due to stretching of staples, which is consistent with the measured curvature within the 37\% uncertainty on the model.  The blue dashed line shows the best quadratic fit. }
\label{fig:staplepolyfit}
\end{figure}

%Tamar's reported stress measurements: The samples consisted of 100g of staples, with a diameter measured as 40.6 mm and approximately 1.5 times the diameter of the cylindrical base.
A stress-strain relation was measured for triaxial compression experiments of packings of staples repeated three times, in each case starting with a new batch of staples, as some staples were permanently plastically deformed by the compression.   An average of the three stress-strain curves is shown in Fig.~\ref{fig:staplepolyfit}, with a range representing the standard deviation of the three curves.  Tests of other staples with the same backbone length and longer arms of 14 mm long or 19 mm long showed qualitatively similar strain stiffening (not shown here).  

 The staples enclose an area that corresponds to a free enclosed area $A$ of 80 times the arm or backbone cross-section $wt$ (see Fig.~\ref{fig:chains}c), which is large compared to the longest chains we measured in Fig.~\ref{fig:entanglefreeprojareawangle}, so a large number of entanglements is expected.   Likewise, simulations of staple packings for the aspect ratio $L_x/L_y=0.37$ produced an average of 6 entanglements per particle \cite{GFHG12}, more than enough for system-filling clusters according to the Erd{\"o}s-R{\'e}nyi model. 	Our observation of strain-stiffening staple packings is consistent with our  observation for chains in Sec.~\ref{sec:area} that average effective free enclosed areas that  are large enough to entangle more than one particle result to strain-stiffening behavior. Since the number of entanglements as a function of arm length has already been studied \cite{GFHG12},  here we focus on identifying the source of stress and extending the model for the stress-strain relation to staples, which is less straightforward.

\subsection{Particle bending observations}
	
\begin{figure}
\centerline{\includegraphics[width=2.5in]{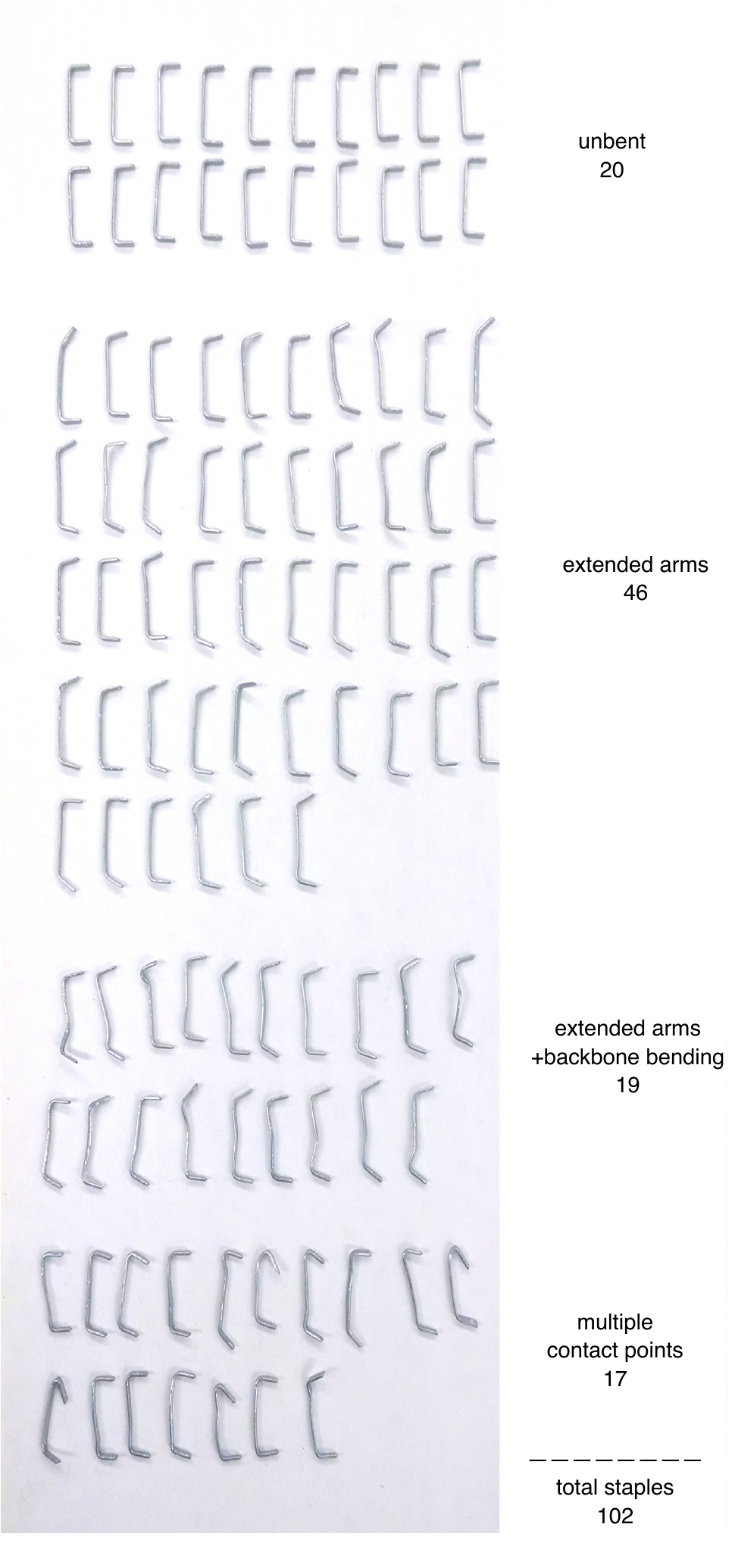}}
\caption{Plastically bent staples from a strain stiffening sample after triaxial compression.  Shadows can be seen from two point-sources of light, which allows visualization of some of the bending in the third dimension.  The majority of staples experienced extensional forces on the arms, indicating extensional forces from entanglement were significant.  This confirms that entanglement and tension on particles plays an important role in strain stiffening for additional particle shapes.
 }
\label{fig:staplebends}
\end{figure}
%trial 4
%images of bent staples
To identify the source of stress, we look at plastically deformed staples after triaxial compression of a packing.  Figure \ref{fig:staplebends} shows a random sampling of 102 plastically bent staples that were poured out and sorted after a triaxial compression test with a maximum strain of 47\%.   The bends can be categorized based on where contact forces would have to be to produce the bends.	20\% of staples showed no resolvable plastic deformation, indicating that the stresses on those particles were below the yield stress for plastic deformation.      45\% of staples have the arms bent outward, suggesting they were pulled apart due to extensional loadings on both arms.   Another 19\% of staples had arms bent outward with an additional inward bend in the backbone.  This bend of the backbone could be due to bending from tensile forces on the arms if the contact points are offset from the backbone, or the backbone could be bent due to an additional contact pushing directly on the backbone along with tensional forces on each arm.   Combined, this is a total of 64\% of the staples that are clearly under some sort of tension.  17\% of staples had deformations that indicate a mixture of extensional forces on one arm and compression loadings on the other arm, corresponding to a more complex force balance requiring multiple contact points for mechanical equilibrium which are not apparent from the plastic deformation alone.   Notably, no staples were bent in such a way to indicate compressive forces on both arms, even though the sample overall was under compression.    Overall, the vast majority of deformed staples experienced extensional forces on the arms, indicating extensional forces from entangling were dominant.  This is consistent with  previous simulations of packings that found tensile stress on S-shaped and U-shaped particles \cite{KMMRA22}, as well as expectations based on the observations of chains in Sec.~\ref{sec:linklength} that tension on particles plays an important role in strain stiffening.

\subsection{Model for stress}

%model conversion for staples
%These two parameters were obtained from x-ray tomography measurements for chains.  Since rigid staples have permanent plastic deformation, we have the advantage that we can determine the probability of staples bending, and the distribution of particle strain by looking at individual particles such as in Fig.~\ref{fig:staplebends}, without need for in-situ imaging.   
The model of Eq.~\ref{eqn:dtaudgammastretch}  for the strength of entangled packings expresses the slope of the stress-strain curve in terms of the probability of particles being stretched $p_s$, and the ratio of local strain to global strain $\langle\partial\gamma_p/\partial\gamma\rangle$.  Since the plastic deformation of staples is more easily expressed in terms of a permanent plastic deformation angle of the bent arms $\theta_p$ instead of the linear strain of a chain $\gamma_p$, we rewrite the chain rule of Eq.~\ref{eqn:dtaudgamma} in terms of the arm bend angle $\theta_p$.  While the load on particles is applied in the direction of tension, the bending in Fig.~\ref{fig:staplebends} suggests that the internal stress is mainly bending stress at the corners of the staples, and perhaps in the backbone. 
%.  This assumes that different components of stress have similar magnitudes, as would be expected from stress transformations (stress components are within a factor of 2 at the same point in a material), 
For extensional forces $F_p$ applied to the arms at a lever arm distance  $l$ away from the corners of the staple, the average stress at the surface of the staples due to bending is approximately $\sigma_p = 6lF_p L_y/wt^2(L_y+2L_x)$ (see Appendix \ref{sec:bendingstresscalc} for derivation of relationship between normal stress and force). 
%theta should be proportional to bending stress, not F
Thus, an appropriate application of Eq.~\ref{eqn:dtaudgamma} and \ref{eqn:dtaudgammastretch} to staples is
%\be
%\frac{\partial \sigma}{\partial \gamma} = \left<\frac{\partial \sigma}{\partial \sigma_p}\right> \left<\frac{\partial \sigma_p}{\partial \theta_p}\right> \left<\frac{\partial \theta_p}{\partial \gamma}\right> = \phi p_s(\gamma)\frac{\partial F}{\partial\theta_p} \frac{3l}{wt^2}  \left<\frac{\partial \theta_p}{\partial \gamma}\right> \ .
%\label{eqn:dtaudgammabend}
%\ee
%\begin{multline}
%\frac{\partial \sigma}{\partial \gamma} = \left<\frac{\partial \sigma}{\partial \sigma_p}\right> \left<\frac{\partial \sigma_p}{\partial \theta_p}\right> \left<\frac{\partial \theta_p}{\partial \gamma}\right> \\
%= \frac{\phi p_s(\gamma)}{2}\frac{\partial F_p}{\partial\theta_p} \frac{3l+t}{wt^2}  \left<\frac{\partial \theta_p}{\partial \gamma}\right> \ .
%\label{eqn:dtaudgammabend}
%\end{multline}
\begin{multline}
\frac{\partial \sigma}{\partial \gamma} = \left<\frac{\partial \sigma}{\partial \sigma_p}\right> \left<\frac{\partial \sigma_p}{\partial \theta_p}\right> \left<\frac{\partial \theta_p}{\partial \gamma}\right> \\
= \phi p_s(\gamma)\frac{\partial F_p}{\partial\theta_p} \frac{6lL_y}{wt^2(L_y+2L_x)}  \left<\frac{\partial \theta_p}{\partial \gamma}\right> \ .
\label{eqn:dtaudgammabend}
\end{multline}
The slope $\partial F_p/\partial \theta_p$ relating the tensile force on a staple $F_p$ to the plastic deformation angle $\theta_p$ of its arms is measured in Appendix \ref{sec:force_theta} analogous to the measurement of $E_p$ for chains.

\subsubsection{Probability of stretching and local deformation ratio}

%probability of bending
\begin{figure}
\centerline{\includegraphics[width=2.7in]{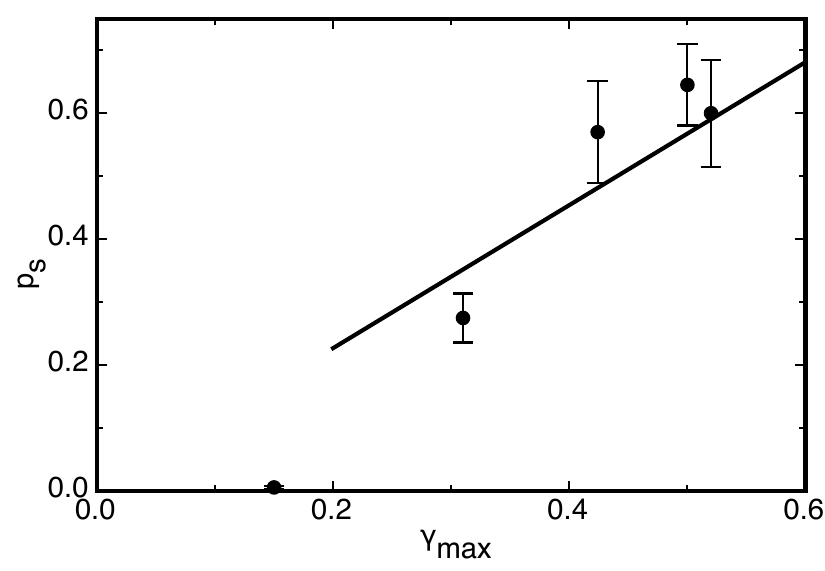}}
\caption{Probability $p_s$ of staples being plastically deformed so as to indicate tensional loads on particles, as a function of maximum packing strain $\gamma_{max}$.  At small strain $\gamma_{max}$ the typical loads may be too small to cause permanent plastic deformation.  The increasing trend $p_s(\gamma_{max})$ is predicted by the model to result in strain stiffening.
Line: proportional fit for $\gamma_{max}>0.2$.}
\label{fig:bendprob}
\end{figure}

To identify the parameters $p_s$ and $\langle\partial\theta_p/\partial\gamma\rangle$ for staples, we sampled particles after experiments with different maximum  triaxial strain $\gamma_{max}$ applied to a packing, as in Fig.~\ref{fig:staplebends}.  By reporting bend angles based on the final strain $\gamma_{max}$, this assumes that the plastic deformation is a function mainly of the maximum strain $\gamma_{max}$ and not highly dependent on the history of slowly increasing strain.

The probability of stretching $p_s$ is measured as the fraction of staples in the categories with extended arms (with and without backbone bending) based on the categorization in Fig.~\ref{fig:staplebends}. This probability $p_s$ is plotted as a function of the maximum strain $\gamma_{max}$ of each experiment in Fig.~\ref{fig:bendprob}, with error bars representing (Poisson) counting statistics.   For small $\gamma_{max} \le 0.15$, we find  almost zero deformed particles.   %The smaller$\gamma_{max} \le 0.15$, %this corresponds to the range where the total load is less less than 3800 N. %22kN at max strain
%using Xuan's counts from Tammar's compression measurements, as those have self-consistent counting of $p_s$ and $\langle\theta\rangle$
In this range, it is likely that most bent particles had forces on them less than the yield force or elastic deformation limit (about 9 N from Fig.~\ref{fig:torque_angle} in Appendix \ref{sec:force_theta}),  and so do not produce permanent deformation.  Thus, we are unable to infer stress distributions from plastic deformation of particles in this range.  For larger $\gamma_{max} > 0.2$ mm, we find significant fractions of bent particles.  To obtain a coefficient for the model trend $p_s(\gamma)$ assuming it is linear, we fit a proportionality coefficient for $\gamma_{max} \ge 0.2$, resulting in $p_s = (1.13\pm 0.07)\gamma$. %conssitency with reasonable chi^2, but too make assumptions about missing data to draw conclusion that confirms linearity
  This may underestimate the number of stretched particles if there are a significant number with forces below the elastic deformation limit.  For the sake of applying a lowest-order model, we will assume that the linear proportionality of $p_s(\gamma)$ extends to the small-$\gamma$ range to account for particles whose forces were below the elastic deformation limit.  
%Note that even though there is some judgment in terms of how small a plastic deformation we can report as resolved, a bias that increases $p_s$ will reduce $\langle\theta\rangle$, so they will mostly cancel each other out

%mean bend angle measurements
\begin{figure}
\centerline{\includegraphics[width=2.75in]{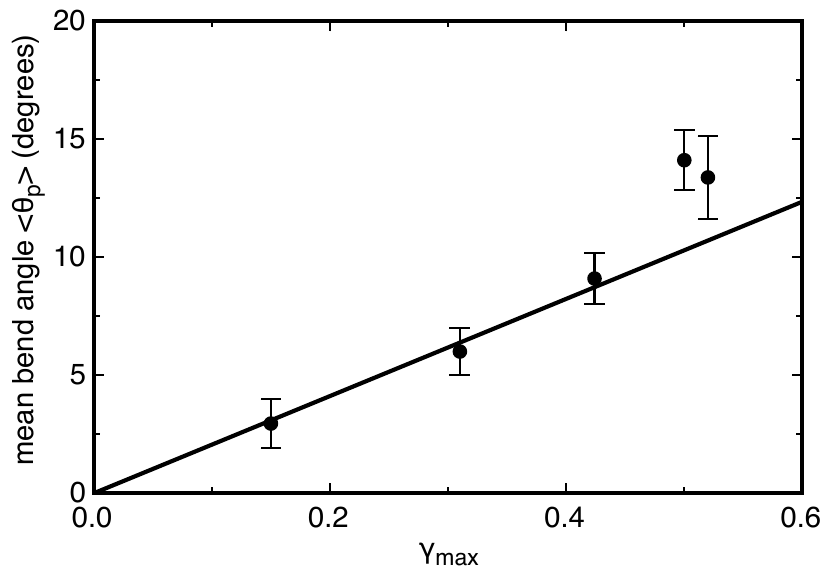}}
\caption{Mean permanent deformation angle $\langle\theta_p\rangle$ of staples that showed extensional plastic deformation after packings were strained to $\gamma_{max}$.  The slope is used to represent the model parameter $\langle \partial\theta_p/\partial\gamma\rangle$. 
}
\label{fig:staplebendangle}
\end{figure}

To obtain $\langle\partial\theta_p/\partial\gamma\rangle$, we measure the average staple bend angle $\langle\theta_p\rangle$ due to plastic deformation among the staples with extended arms that we counted as bent for the purposes of calculating $p_s$. Figure \ref{fig:staplebendangle} shows  the mean bend angle $\langle\theta_p\rangle$ of bent staple arms as a function of maximum strain $\gamma_{max}$.  The error bars represent  the standard error of the bend angles of measured particles.  A linear fit to the data for $\theta_p < 11^{\circ}$ (the same range where the load $F_p$ is linearly fit to $\theta_p$ in Fig.~\ref{fig:torque_angle} of Appendix \ref{sec:force_theta}) yields $\partial \theta_p/\partial \gamma_{max} = (20.6\pm1.9)^{\circ}$, with the linear fit consistent with the data within the standard errors. The linearity here validates the model assumption that $\langle\partial\theta_p/\partial\gamma\rangle$ is a constant. At larger strain $\gamma_{max}$, the more rapid and less systematic increase in $\langle\theta_p\rangle$ may result from plastic deformation, which occurs for $\theta_p>11^{\circ}$ (see Fig.~\ref{fig:torque_angle})

\subsubsection{Test of the stress-strain relation}

%fit
We now have measured values needed to evaluate Eq.~\ref{eqn:dtaudgammabend} to predict the stress-strain relation of staple packings. 
Using $\partial F_p/\partial\theta_p =  (6.5\pm0.9)$ N/$^{\circ}$ and $l=0.6\pm0.2$ mm from Appendix \ref{sec:force_theta}, 
$w=1.24$ mm, $t=0.48$ mm,  $\phi=0.182$, $p_s(\gamma) = (1.13\pm 0.07)\gamma$ from Fig.~\ref{fig:bendprob}, and $\partial \theta_p/\partial \gamma_{max} = (20.6\pm 1.9)^{\circ}$ from Fig.~\ref{fig:staplebendangle} in Eq.~\ref{eqn:dtaudgammabend} results in $\partial \sigma/\partial \gamma = (198\pm 74)\gamma$  MPa, which integrates to 
\begin{equation}
\sigma = (99\pm37)\gamma^2 \mbox{ MPa} \ .
\label{eqn:modelstaples}
\end{equation}
For comparison, a quadratic fit of the stress-strain curve for staples yields $\sigma = (62.76 \pm 0.06)\gamma^2$ MPa, shown in Fig.~\ref{fig:staplepolyfit}.   This fit has a coefficient of determination of 0.99, corresponding to a standard deviation of 0.5 MPa around the fit, indicating that the quadratic scaling is a good approximation.   The coefficient for the model curvature is consistent with the fit value within the 37\% model uncertainty.    Thus, we can attribute the stress-strain relation of packings of staples to the deformation of particles due to tensile loadings.  

%We used the approximations of a constant $\partial F_p/\partial\theta_p$ and linear $p_s(\gamma)$ for the sake of a simple lowest-order model which captures the dominant quadratic term in Fig.~\ref{fig:staplepolyfit}.  The non-linear nature of $\partial F_p/\partial\theta_p$ (see Appendix \ref{sec:force_theta}), the lack of plastic deformation to resolve $p_s$ at small $\gamma_{max}$, as well as the interaction with the elastic boundary could be hiding a more nuanced stress-strain relationship at small strain.

 For completeness, we tested a quadratic fit with two fit parameters which yields $a_1=(2.71\pm0.04)$ MPa and $a_2 = (59.4\pm0.2)$ MPa.  In this case, the coefficient $a_2$ only changes by a few percent, and $a_1$ only contributes to $9\%$ of the total stress at $\gamma=0.5$. Thus, the strength of packings of entangling staples can also be approximated by a dominant quadratric function without a linear term, which is again attributed to a probability of entanglement $p_s$ proportional to increasing strain $\gamma$, similar to chains.  With the linear term, the data fits with a reduced $\chi^2=1.0$ when using the standard deviation of multiple runs as an uncertainty, indicating the linear term is reproducible in the case of staples, if relatively small. 
% $p_s$ might be underestimated by only counting plastic deformation -> offset?

\section{Predictions of $A_{eff}/A$ for rigid particles} %and the probability of stretching $p_s$ for rigid particles}
\label{sec:ps_prediction}

We have presented a model to predict the number of entanglements per enclosure $n$ which determines whether there are system-filling clusters of entanglements to support strong-strain stiffening.  %These models required as input the probability of stretching $p_s$, and the local strain ratio $\partial\gamma_p/\partial\gamma$ or $\partial\theta_p/\partial\gamma$,
This model required as input the free entangled area $A_{eff}$.   This could be obtained from either {\it in situ} imaging (e.g.~x-ray tomography) or computer simulations. % or from a post-experiment analysis of some indicator of contact or stress (e.g.~plastic deformation), which will be limited to certain types of particles or experimental techniques. % ($E_p$ and $\phi$ are relatively easy to measure).  
Alternatively, we desire a simple modeling approach.  

%prediction of $A_{eff}$
 \begin{figure}
\includegraphics[width=2in]{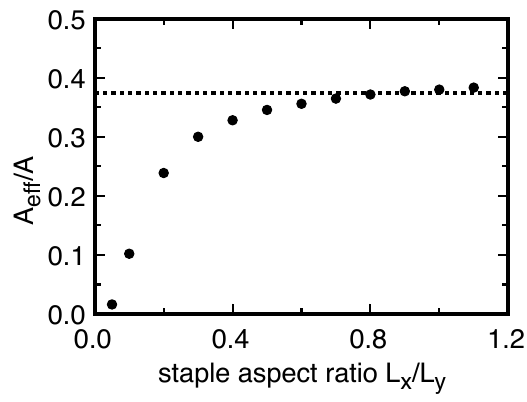}
\caption{The ratio $A_{eff}/A$ accounting for excluded volume effects in three dimensions for staples of different aspect ratios. Dotted line: result obtained from fitting number of entanglements $n$ from \cite{GFHG12}.  The value is nearly constant for staple arms much longer than the arm width, corresponding to the limit used in the entanglement density model of \cite{GFHG12}. 
}
\label{fig:Aeffratiostaples}
\end{figure}
%consistency of A_eff/A for staples
 For staples, the number of entanglements per enclosure $n$ has been  obtained  previously using the same entanglement volume model from Sec.~\ref{sec:area}, which  is equivalent to $n=\phi A_{eff}/tw$  \cite{GFHG12}, so we need to predict $A_{eff}$.   Appendix \ref{sec:Aeffcalc} shows the details of calculating $A_{eff}/A$ for the rigid geometry of staples using a Monte Carlo simulation considering the possible positions and orientations of an entangled staple arm.  The resulting values of $A_{eff}/A$ for staples are shown in Fig.~\ref{fig:Aeffratiostaples} for different staple aspect ratios $L_x/L_y$ with a fixed backbone length $L_y = 13a$, corresponding to the simulation of \cite{GFHG12}.   The range of values of $A_{eff}/A$ overlaps with the best fit $1/\alpha=0.378\pm 0.016$ from \cite{GFHG12}.   The plateau in the limit where the width of the entangled region is much larger than the particle width (aspect ratio $L_x/L_y \gg 1/13$) is consistent with their argument that this parameter is nearly a constant \cite{GFHG12}.
%For our aspect ratio 0.36, we find $A_{eff}/A=0.421$.  This is only 11\% larger than the fit value $1/\alpha=0.378\pm 0.016$ from \cite{GFHG12}, 
 $A_{eff}/A$  does drop off  for small aspect ratios  due to increased excluded volume effects on the possible angles $\Theta$ of entangled particles relative to the entangled plane as the width of the entangled region becomes comparable to the staple thickness.  A similar trend is seen in Appendix \ref{sec:Aeffcalc} for chains based on x-ray tomography data.  The increase in $A_{eff}/A$ towards a plateau as the entangled region becomes wide compared to the particle thickness should be a general feature regardless of particle shape.  In general, the maximum limit value of $A_{eff}/A$ is 0.5 for particles with very thin arms and large entangled areas (see Appendix \ref{sec:Aeffcalc}). With these models, $A_{eff}$ and $n$ can be calculated from simple Monte Carlo simulations for arbitrary rigid shapes to predict whether the particle will form system-filling clusters ($n\stackrel{>}{_\sim}1$) and exhibit strong strain stiffening.

\section{Discussion}

%strength of entanglement-based grasper scales with pressure/modulus of entangling strands \cite{BTCW22} %Active entanglement enables stochastic, topological grasping

%Jammed systems have a power law scaling of contact number in packing fraction or compressive strain \cite{OLLN02}  or F. Bolton and D. Weaire, Phys. Rev. Lett. 65, 3449 (1990).
%... jammed systems also have a gradual increase in contacts with compression with increasing load per contact, such that Hertzian contact models resulting in strain stiffening.

%strength limited by deformation
How do the mechanics of entangled structures relate to other granular systems? While soil strength is usually limited by the confining pressure or interparticle attractions, we find that in packings where entanglements fill the whole system the collective strength scales with link strength.  This limit is similar to textiles, chain links, and other permanently entangled patterned structures \cite{WLHAD21}, as in these cases the structural constraints prevent particles from shearing past each other to rearrange, forcing  links to deform in response to stress.  However, those patterned structures do not often have significant strain stiffening.  We have argued that strain stiffening comes from the increasing number of stretched links with increasing strain, which is not a gradual increase in patterned systems, which instead have an initially linear stress-strain relation starting when all of the links touch at about the same time.

 %significance of entanglement 
Why should entanglement rather than other types of contacts found in granular packings be the constraint relevant to strong strain-stiffening? While frictional contacts between particles provide constraints on motion, the force from those frictional constraints is limited by a confining stress, which is the typical limitation of strength in soils and packings of convex particles.   In contrast, the concave part of the geometry of an entangling particle allows for the entangled particle to be potentially stable when pushing against the entangling particle, providing a stronger constraint against motion towards the entangling particle, assuming there are enough constraints to prevent pushing the entangling particle out of the way such as  through rearrangements from a shear band failure.   Shear banding can be prevented by the extra mechanical constraints provided by a long particle perpendicular to the potential shear band \cite{RR13}, so a system-filling structure of entanglements prevents shear bands along any plane.  With these constraints, further strain of the system requires particle deformation, which results in the particle deformation determining the stress.   Our definition of entanglement is a simple binary parameter that identifies some mechanical stability as a constraint on a component of motion of entangled particles toward the entangling particle.  However, this is not a detailed accounting for this stability against all components of motion.   It remains to be seen if this definition of entanglement is a sufficiently general algorithm to identify systems that strain stiffen, or if marginal cases might require a more complicated accounting of the mechanical constraints.

 %rods
It remains an open question of whether this model could be generalized to strain stiffening packings of rod-shaped particles \cite{BWBKGK22}.  Since straight rods entangle no volume, they cannot entangle other particles.  However, if the rods bend into arcs under small loads, they could in principle entangle the particles that are pushing them to bend according to our definition of entanglement based on free enclosed area. For bent rods, any stress due to entanglement would come strictly from bending of particles due to multiple points of loading, rather than any tensional load on the particles, suggesting the important aspect of the loading is not necessarily the tension on particles, but more generally that the loading produce stress in response to deformation of particles. %In the case of rods, positive curvature of the stress-strain curve may also come from an increasing number in contact points, which decreases the distance between contact points and increases the effective stiffness \cite{BWBKGK22}.

 %small fraction of fiber reinforcement 
 It is interesting to consider if this understanding of entanglement could lead to mixtures with enhanced strength or strain stiffening in soils, concrete, and other granular materials.
Fibers are typically only added to soils at a fraction less than about 1\% to provide reinforcement, and while this reinforcement increases strength, it does not produce strain-stiffening when the fibers are mixed in randomly \cite{AssadiLangroudietal22}.   The small fraction of fibers is likely too little to produce strong strain stiffening from system-filling clusters of entanglements.  The same could be said about concrete reinforced with fibers \cite{AHFW92}.   Larger packing fractions of fibers have been confirmed to enhance the weaker self-amplified friction mechanism for strain stiffening from localized strain \cite{DHRSPD18}.
%WHY IS THIS DIFFERENT FROM RR13 -- RR13 in 2D, has shear plane forced perpendicular to chain rather than natural failure plane or random arrangement
%packing fraction of chains relative to spheres was varied by DHRSPD18 for local indentation, only measured up to 17 kPa averaged over sample cross section, 750kPa on indenter

%comparison to MacKintosh
Another model of for athermal fibrous gels produces strong strain stiffening \cite{SLRVJKM16}.  That system is modeled as an arrangement of particles with elastic bonds with tension and bending stiffness between some fraction of neighbors.  Under strain, rearrangements are mostly prevented by the bonds, so increasing strain on the system increases the number of stretched bonds, resulting in strain stiffening.  The source of stress is analogous to our case, where the increasing stretching results in strain stiffening, although in our case it is physical entanglements that prevents rearrangement and cause more entangled particles to be deformed instead of gel bonds. The fiber model requires a numerical simulation, in contrast to our model which breaks down the stress contributions into a low-dimensional model using Eq.~\ref{eqn:dtaudgammastretch}. In the fiber model, the fraction of bonds with neighbors is assumed to be enough that the system would be above the gel percolation threshold, which would correspond to greater than one connection per node in the Erdos-Renyi model.  This parameter space is mathematically analogous to our experiments with system filling clusters of entanglements ($N > 8$).  The fiber model emphasizes that the strain stiffening is a critical behavior with a critical transition in strain, which is different from the critical transition we find with increasing $N$.  Our system does not have such a critical transition in strain, as the contribution of the elastic membrane to the stress-strain curve is dominant at small strains. More generally,  in other macroscopic particle granular systems the small-strain region is usually dominated by a confining stress which is required to prevent the system from collapse due to its self-weight (even highly entangled systems tend to have a partial collapse without some confinement).

\section{Conclusions}

In this paper, we presented a model that can quantitatively describe the stress-strain relation of strong strain stiffening due to stretching of entangled particles in random packings,  and predictions that determine whether particles of known shape will form the system-filling clusters of entanglements required for this strong strain stiffening.  

%system filling clusters
Using x-ray tomography measurements of the packing structure, we showed that the number of entanglements $n$ of chains is nearly proportional to an effective free enclosed area $A_{eff}$ surrounded by a chain available to entangle other particles, accounting for excluded volume effects (Fig.~\ref{fig:entanglefreeprojareawangle}). This area $A_{eff}$ and the expected number of entanglements $n$ can be calculated given the shape of the particles in the packing assuming random particle arrangements (Fig.~\ref{fig:freeprojarea}). The threshold of about 1 entanglement per enclosure required to produce system-filling clusters and strong strain stiffening is similar to the Erd{\"o}s-R{\'e}nyi model for random graphs, which is often used to describe the gelation transition.   

%generality of predictions
We presented this model in a form that could be extended to packings of different particle shapes that can entangle.  In particular, for a particular rigid particle geometry, one could calculate from a Monte Carlo simulation the possible entanglement arrangements the effective free enclosed area $A_{eff}$ (Appendix \ref{sec:Aeffcalc}), which leads to the number of entanglements $n$, and thus whether or not there are enough entanglements to connect in system-filling clusters to produce strong strain stiffening.  While these quantities can be predicted from the known geometry of rigid particles (Fig.~\ref{fig:Aeffratiostaples}), calculating these quantities required detailed structural measurements for flexible chains to calculate $A_{eff}$ and $n$.

%force network and strength
For both chains and staples,  axial compression of the packing leads to tensional loads on particles (Figs.~\ref{fig:linkdist}, \ref{fig:staplebends}), which suggests the strength of the packing comes from the stiffness of particles as they are stretched.   The stress supported by the stretched particles is orders of magnitude higher than the confining stress from the membrane at the side of the packing, which would usually limit the strength of granular packings.   These large forces from stretching particles are only found in packings of chains long enough to produce system-filling clusters of entanglements \cite{BNAJ12}, which allows the large forces to be supported along a force network that spans across the entire packing.

%stress model
We presented a model for the stress-strain relation of random packings with system-filling entanglements, where the stress originates from the stress required to deform the particles, which is due to tensional loads on the particles in the cases tested.  Using a mean-field model, the slope of the stress-strain curve of the packing was written as a product of an effective particle modulus, the packing fraction, the probability of stretched particles $p_s$, and the ratio of particle strain to packing strain (Eq.~\ref{eqn:dtaudgammastretch}).  For both chains and staples, the increase in effective stiffness with strain that characterizes strain stiffening can be attributed to the increase in the fraction of stretched particles $p_s(\gamma)$ as packing strain $\gamma$ increases (Fig.~\ref{fig:linkdist}b, \ref{fig:bendprob}) and causes more entangled particles to pull on each other, while the effective particle stiffness and the ratio of particle strain to packing strain are approximately constant (Figs. \ref{fig:chainextension}, \ref{fig:torque_angle}, \ref{fig:staplebendangle}).  A model with a linear increase in the probability of stretched particles $p_s$ with packing strain $\gamma$ corresponds to a quadratic stress-strain relation, which is a decent approximation of the observed stress-strain curves.  The curvature of the stress-strain relation was predicted within the model uncertainties ($\le 37\%$) of measured values for packings of both chains and staples (Figs.~\ref{fig:stressstrain}, \ref{fig:staplepolyfit}).   To predict the stress-strain relation, the effective particle modulus and packing fraction are relatively easy to measure, however, the probability of stretching and the ratio of particle strain to packing strain required information from experiments with three-dimensional imaging, permanent plastic deformation, or simulations of packings, and are not yet easily predictable from first principles.
% the probability of stretching could be predicted from $A_{eff}$.  

%system size
%We observed strain stiffening in cases where the chain length is about half of the system size, so that it is reasonably likely any chain can interact with any other chain in the system.  Entanglement statistics and parameters such as the critical chain length for strain stiffening may change for larger system sizes compared to the chain length.  %Understanding whether strain stiffening is less likely in larger systems would be relevant to understanding systems like soil strength due to roots, ...

%\subsection{Open questions}

%comparison to self-amplified friction
A comparison in Appendix \ref{sec:expmodel}  indicates that the strain stiffening observed in previous experiments with a localized indenter resulting in an exponential stress-strain relationship due to self-amplified friction is a different mechanism for strain stiffening than what we observed in triaxial compression experiments \cite{DHRSPD18}.  This indicates that neither strain stiffening mechanism is an intrinsic material property, but rather the behavior depends on the boundary conditions as well.  In our case, the boundary confinement provided by the wide compression area and elastic sidewall presumably helps the system-filling entangled structure support a force network all the way across the system, while a localized indenter with a more open system allows more localized shear that instead activates self-amplified friction \cite{DHRSPD18}.  Whether other entangling systems like  disorganized matted fibers strain stiffen because of self-amplified friction or an increasing fraction of stretching of fibers, or whether the behavior generally depends on how localized the strain is, remain open questions, but we now have identified methods which can help answer these questions for other systems.

%outcomes
%... extension to roots in soils, 
%A simple approach for estimating contribution of vetiver roots in shear strength of a soil–root system, Faria Fahim Badhon, Mohammad Shariful Islam, Md. Azijul Islam & Md. Zia Uddin Arif :shear strenght linear in tensile strength of roots
%...Since strength is a key design parameter, this modeling can aid in determining important parameters for designing optimal shapes for pourable construction materials.

\section{Acknowledgements}

We thank Mike Dunlap for assistance with materials testing equipment, Peter Eng and Mark Rivers for their assistance with x-ray tomography at the Advanced Photon Source, Ling-Nan Zhou for his particle tracking code, Nicholas Rodenberg and Dylan Murphy for preliminary measurements, Sulimon Sattari for performing simulations of network models for comparison, %not used
Henos Musie for performing measurements for Fig.~\ref{fig:torque_angle},  Tammer Abiyu for performing measurements for Fig.~\ref{fig:staplepolyfit}, David Brantley for performing measurements for Fig.~\ref{fig:chainbending}, and Xuan-Truc Nguyen for analyzing particles for Figs.~\ref{fig:staplebends} and \ref{fig:staplebendangle}.   We acknowledge the Advanced Photon Source general user program at Argonne National Laboratory for  providing beam time.  This work was supported by the NSF MRSEC program under DMR-0820054 and DMR-2011854.

A. Nasto and A. Athanassiadis performed x-ray tomography measurements, particle tracking, and chain identification.  A. Nasto did experiments for data in Fig.~\ref{fig:stressstrain}. E. Brown did the remaining analysis, modeling, and wrote the draft.  E. Brown and H. Jaeger planned experiments.  E. Brown and K. Mitchell discussed modeling approaches.

\appendix

\section{Self-amplified friction model comparison}
\label{sec:expmodel}

A model for the stress-strain relations for packings of chains based on self-amplified friction and motivated by polymer models was presented by \cite{DHRSPD18}.  Their stress-strain data fit well  to exponential functions.  This was explained to be due to self-amplified friction, where friction contributes to the normal force that determines the frictional force as beads move around locking points in the structure similar to a wormlike chain model.    The resulting stress-strain relation is
%$F \propto \exp(\mu\sqrt{N}z/b)$
\begin{equation}
\sigma = \sigma_0\exp\left(\Lambda\gamma\right)
\label{eqn:expfit}
\end{equation}
The characteristic inverse strain scale $\Lambda$ is predicted to be proportional to the number of locking points that the chain could get caught on,  and assuming that chains arrange statistically similar to thermal polymer chains so the radius of gyration or end-to-end distance proportional to $\sqrt{N}$ results $\Lambda \propto \sqrt{N}$ \cite{DHRSPD18}.  They observed both the exponential scaling of stress with strain, and the dependence of $\Lambda \propto \sqrt{N}$. However, their observation contrasts with our claims of strong strain stiffening only appearing above a minimum chain length $N\ge 9$  \cite{BNAJ12}, with a sharp (rather than continuous) dependence on $N$.  
%where $L=24$ is the height of the cell in units of bead diameters.  

To test their model against our data, we fit an exponential $\sigma(\gamma)$ to the data in Fig.~\ref{fig:stressstrain} for $0.07 <\gamma<0.2$, since for $\gamma<0.07$, the stress-strain curves are concave down (opposite curvature of an exponential function), due to the confining stress of the membrane. %We also ignore the limit of packings of spheres ($N=1$), as the stress increases nearly linearly with strain for small strain before leveling off at a yield stress once the packing has dilated to the critical state.
\begin{figure}
\includegraphics[width=3.in]{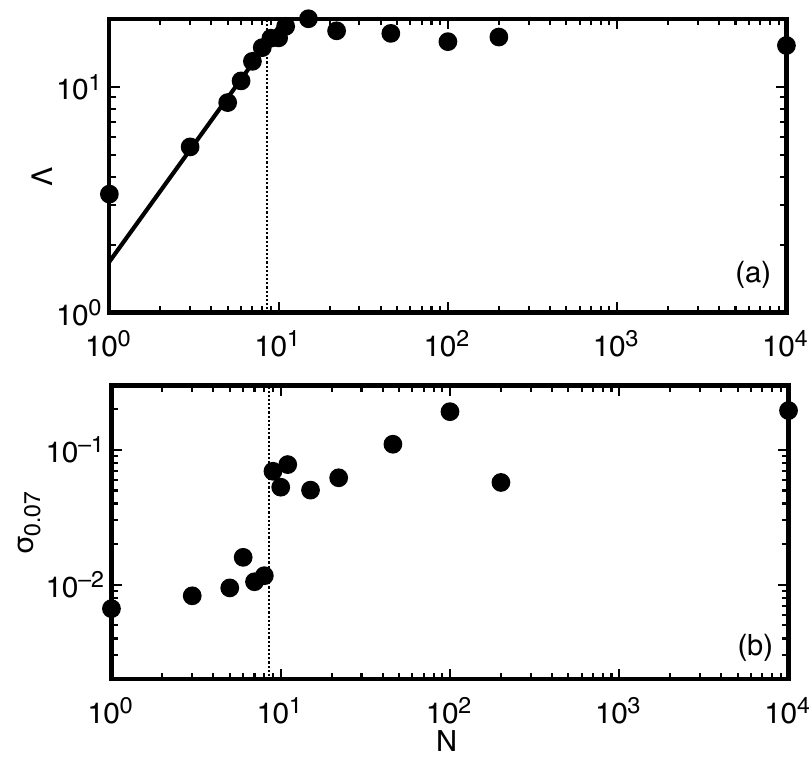}
\caption{Parameters from a fit of the stress-strain curve to an  exponential function (Eq.~\ref{eqn:expfit}). (a) exponential coefficient $\Lambda$.  (b) the stress $\sigma_{0.07}$ at $\gamma=0.07$.  Both parameters show a sharp transition at the onset of strong strain stiffening at $N=9$.  $\Lambda$ does not scale as the wormlike chain model prediction of $N^{0.5}$ in either regime, and the sharp transition in $\sigma_{0.07}$ is not captured by the wormlike chain model.  The corresponding model of self-amplified friction does not explain our observed strong strain stiffening under triaxial compression in a confined geometry.
}
\label{fig:expfit}
\end{figure}
A plot of the fit parameter $\Lambda$ is shown in Fig.~\ref{fig:expfit}a.  A fit yields $\Lambda \propto N^{1.04\pm0.05}$ for $3 \le N \le 8$, and $\Lambda \propto N^{-0.02\pm 0.12}$ for $N \ge 9$.  This has a distinct change in scaling after the transition to strong strain stiffening, however, neither range has the exponent $0.5$ predicted from random-walk statistics of a thermal chain in the wormlike chain model\cite{DHRSPD18}.   Since $\sigma_0$ corresponds to the stress value at $\gamma=0$, extrapolated from the fit range, we instead plot the stress at the start of the fit range $\sigma_{0.07} = \sigma_0\exp(0.07\Lambda)$ in Fig.~\ref{fig:expfit}b.     The coefficient $\sigma_{0.07}$ also jumps sharply at the threshold of strong strain stiffening.  Such a discontinuous jump is not expected in the self-amplified friction model based on continuous scalings of $\Lambda(N)$. 

To determine which functions fit the stress-strain curve better, we compare $\chi^2$ values for exponential and quadratic fits, where the $\chi^2$ is the mean-square difference between the data and fit normalized by the uncertainties which are taken as the standard error of multiple repetitions of the stress-strain curve.  We typically find the quadratic fit of Sec.~\ref{sec:model} with two parameters ($a_1$ and $a_2$) to have a $\chi^2$ smaller than the exponential fit with two parameters by an average factor of 3.5 for fits over the range $0.07 < \gamma < 0.20$.   %If we extend the fit range from $0.01 < \gamma < 0.20$, then $\chi^2$ is smaller by an average factor of 16 for the quadratic fit due to the initial concave-down shape of the stress-strain curves.  
If we extend the fit range from $0.07 < \gamma < 0.40$ (we only have data available for $N=10^4$ in this range), then $\chi^2$ is smaller for the quadratic fit by a factor of 350.  These $\chi^2$ values indicate that the quadratic fit is a better description of our data than an exponential function for any of these fit ranges.

Why is the scaling of our stress-strain curves so different from those of \cite{DHRSPD18}?  An important difference is that  \cite{DHRSPD18} used a narrow indenter with an open top surface that allows dilation of the packing without resistance, which results in a more localized strain so a few chains near the indenter are responsible for most of the displacement.  We confirmed experimentally that using a narrow indenter and free boundary allows beads to move past each other when the packing is indented.   This more localized motion allows the mechanism of self-amplified friction to determine which chains move and thus the response measured by the indenter.  In contrast, 
our experiments used a flat plate that produced a relatively uniform strain in the packing, and a weak confinement from the sides with the elastic membrane.  In this case, the entanglements of long chains prevent particles from shearing past each other, thus preventing the self-amplified friction mechanism, and instead requiring load to be generated by deformation of particles.   This comparison  confirms the importance of preventing shear banding \cite{RR13} in the strong strain stiffening behavior that we observed.  However, it also highlights that the apparent measured stress-strain relation is not necessarily an intrinsic material property, and even different mechanisms can exist for strain stiffening for the same material depending on whether the boundary conditions allow a more localized shear banding or prevents it.

%In addition, in our case, the outer membrane stiffness contributes significantly to the stress at small strain, which is in part why we did not fit the exponential function to that range, and it seems likely that the outer membrane was much less relevant for the experiments of \cite{DHRSPD18} to obtain exponential curves even at small strain.  
%...low force?% <100 N, 10^6 Pa on indenter close to ours
%correciton for self similar stress would still be small, scales as $1/(1-\gamma)$ contributes about 1 to $\Lambda$ in our case.
%DHRSPD18 find  increase in force with displacement even for N=1 -> suggests increasing fraction of system participating .  exponential force scaling could come from number of particles that participate in carrying a load is proportional to force  
%Perhaps the localized strain does not lead to system-filling entanglement networks? and exponential model characterizes number of participating chains

\section{Effective extensional modulus of chains}
\label{sec:chainstiffness}

\begin{figure}
\includegraphics[width=2.in]{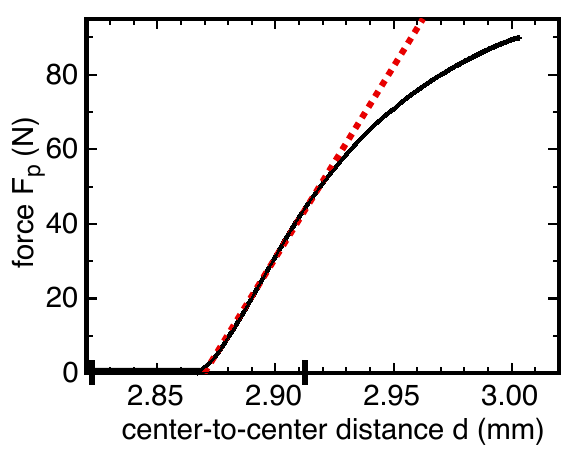}
\caption{Force vs. average center-to-center distance for extension of a single chain.   Dotted line: linear fit to elastic regime.  The curve ends when the chain breaks.  The error bar is the standard deviation in the force onset length $d_0$ from link to link.
}
\label{fig:chainextension}
\end{figure}

To measure the effective extensional modulus $E_p$ of an isolated chain, we used a chain with $\xi=8$ beads and length $N = 98$ beads with the beads at each endpoint held by clamps.  We extended quasistatically at a rate of 0.1 mm/s while measuring the force $F_p$ as a function of the length of the chain between endpoints.  A force curve averaged over five repetitions is plotted in Fig.~\ref{fig:chainextension} against the average center-to-center distance $d$, which is obtained by normalizing the extended chain length by the number of links $N-1$ between the endpoints of the chain.  Because the beads are loosely connected by links that are free to move around up to a point, nearly zero extensional force is measured (within the resolution of 0.01 N) until an average center-to-center distance of $d_0 = 2.864 \pm 0.005$ mm; here the uncertainty $\sigma_{d_0}$ is the standard deviation from the five repetitions.  %There is an additional absolute uncertainty of 0.002 mm on the average center-to-center distance due to the initial length measurement.

An effective stiffness $k$ is obtained by fitting a linear function $F_p = k(d-d_0)$ to the force vs.~center-to-center distance $d$ in the range of force from the resolution limit up to 40 N.  The average slope from the fits is $k = 1050$ N/mm, with a standard deviation of 60 N/mm for the five repetitions.  A little beyond 40 N, the force vs.~center-to-center distance curve becomes non-linear and the deformation is plastic (which we confirmed with a series of tests with a separate chain where we ramped the displacement back and forth over smaller ranges and found non-zero hysteresis loops above 40 N, but not below).  The maximum force reached in the measurement beyond which the chains broke was on average 86 N with a standard deviation of 9 N.

Since the chain has a non-uniform cross-section, the effective chain modulus $E_p$ appropriate to the mean-field model is calculated based on the average cross-sectional area of the bead along the length $z$ of the chain of  $(1/\bar d)\int_{-a/2}^{a/2} \pi (a^2/4-z^2) dz   = \pi a^3/6\bar d$.  This results in  $E_p=6kd_0 \bar d/\pi a^3 = (1500\pm 90)$ MPa using $\bar d=1.11a$  from data in Fig.~\ref{fig:linkdist}.
%in triaxial strain, both stress and strain are 1/2 $\sigma_1$ in radial direction ad stress/strain has constant ratio in all planes, so no effect on calculation

\section{Bend angles of chains}
\label{sec:chainbending}

\begin{figure}
\includegraphics[width=2.in]{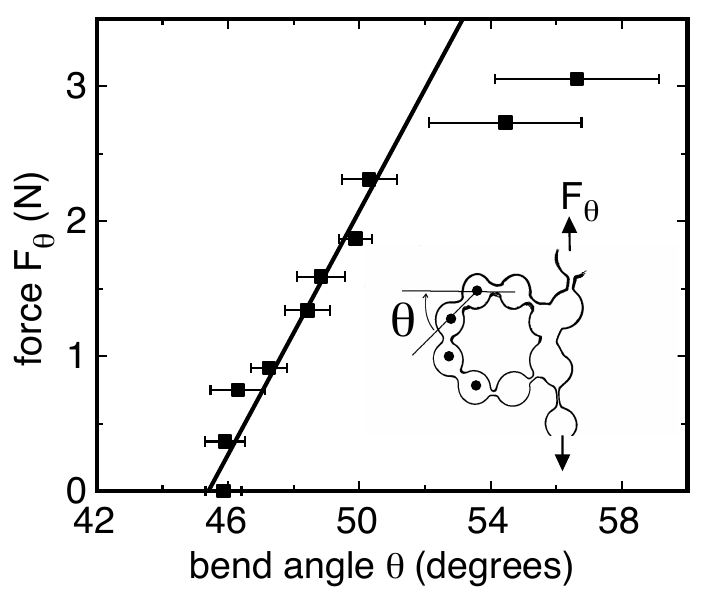}
\caption{Force vs. average bend angle $\theta$ for bending of a single chain.   Solid line: linear fit to elastic regime.  The inset shows the geometry of the bent chain and the bend angle $\theta$.}
\label{fig:chainbending}
\end{figure}

To help determine whether the compression of tight loops of chains contributes to strong strain stiffening, the bending stiffness of a chain loop was measured by pulling on a looped chain.  A chain with $\xi=8$ beads was bent into a loop (shown in the inset of  Fig.~\ref{fig:chainbending}) and held in place between two sheets of acrylic separated by 4.5 mm in such a manner that the chain would not unloop or catch on another bead as the ends of the chain were pulled.  Pictures were taken to measure local bend angles $\theta$ between the links of neighboring chains (shown in the inset of  Fig.~\ref{fig:chainbending}) with an uncertainty of 1 degree limited by the image resolution.  Four angles were measured for each chain (those centered on the beads with solid circles overlaid in their centers in the inset of  Fig.~\ref{fig:chainbending}), and the measurement was repeated for two different chains to obtain a total of eight measured angles at several different extensional force values.  The average angles for each force are shown in Fig.~\ref{fig:chainbending}, with the error bars indicating the standard deviations of the measured angles.  %We use the same force value for each link in a chain, since the measured force must be transferred through each link, assuming that frictional force between the acrylic walls and the chain is negligible.  
We measured no resolvable force until the chains were bent beyond $45\pm 1^{\circ}$ degrees, which corresponds to the minimum loop circumference $\xi =  360^{\circ}/45^{\circ} = 8.0 $. %To obtain an effective bending stiffness $k_{\theta}$ we fit a line to $F_{\theta} = k_{\theta} (\theta-\theta_0)$ for $F_{\theta} < 2.5$ N and obtain a slope $k_{\theta} = 0.45 \pm 0.02$ N/deg and initial force angle $\theta_0 = 45.4  \pm 0.1$ deg.  
While there is an initial linear force $F_{\theta}$ with bend angle $\theta$, above 2.5 N, the chains started to bend non-uniformly, indicated by the large increase in standard deviation of $\theta$ with increasing force shown in Fig.~\ref{fig:chainbending}; in particular one link would bend much more than the others, indicating yielding of the weakest link.

\begin{figure}
\includegraphics[width=3.4in]{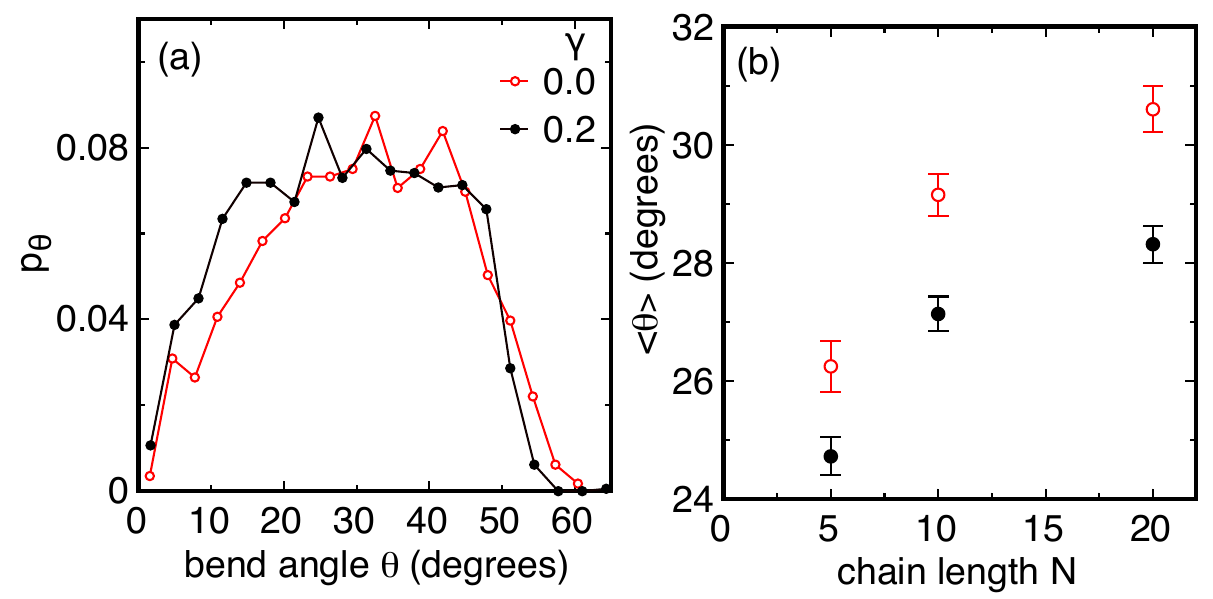}
\caption{(a) The probability distributions of bend angle $\theta$ between neighboring links of chains for $N=20$. Open circles:  $\gamma =0$. Solid circles:  $\gamma=0.2$. (b) The average value  $\langle\theta\rangle$ for each strain $\gamma$ and chain length $N$.  The decrease in $\langle\theta\rangle$ with strain $\gamma$ and the limited bending force inferred from Fig.~\ref{fig:chainbending} at the bend angles in the distribution suggest that the force generated by tightening of chain loops is not significant in strong strain stiffening of random packings of chains.
}
\label{fig:bendangledist}
\end{figure}

To see the significance of the forces due to bending of chains, distributions of bend angles $\theta$ from x-ray tomography measurements of random packings of chains are plotted in Fig.~\ref{fig:bendangledist}a for $N=20$.   The mean value for each data set is plotted in Fig.~\ref{fig:bendangledist}b.  The mean bend angle decreases slightly with strain.  The decrease in the mean value  $\langle\theta\rangle$ indicates that loops are not typically tightening as strain $\gamma$ is applied \cite{LRR11}.  Furthermore, the maximum bend angle for a single link is $65^{\circ}$ at $\gamma=0.2$, $N=20$, corresponding to a maximum force from bending a chain into a loop of only 4 N from extrapolating Fig.~\ref{fig:chainbending}.  This maximum bending force is much weaker than even the average force from extension of chains (of 44 N Sec.~\ref{sec:linklength}), indicating that the contribution to the total stress from compression of loops is negligible compared to that from extension of chains in strong strain stiffening packings.

\begin{figure}
\includegraphics[width=2.7in]{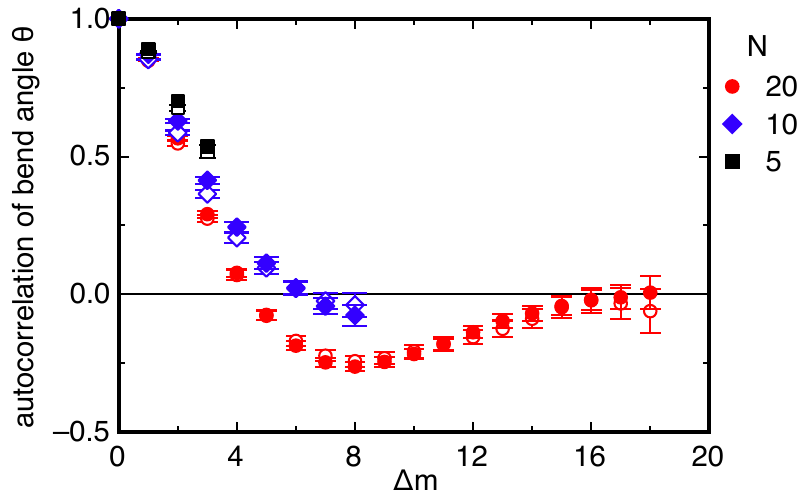}
\caption{The autocorrelation of bend angle $\theta$  as a function of separation by $\Delta m$  beads along a chain.  Circles: $N=20$.  Diamonds: $N=10$.  Squares: $N=5$.  Filled symbols:  strain $\gamma=0.2$.  Open symbols:  strain $\gamma=0$.  The vertical bars indicate a standard error.
}
\label{fig:bendanglecorr}
\end{figure}

To help characterize the shape of chains in packings for Sec.~\ref{sec:clusters},   the autocorrelation of bend angle $\theta$ as a function of separation by $\Delta m$  beads along a chain was calculated from the x-ray tomography data, and  plotted as a function of the separation distance $\Delta m$ beads in Fig.~\ref{fig:bendanglecorr}.

\section{Packing fraction}
\label{sec:phi}

\begin{figure}
\includegraphics[width=3.4in]{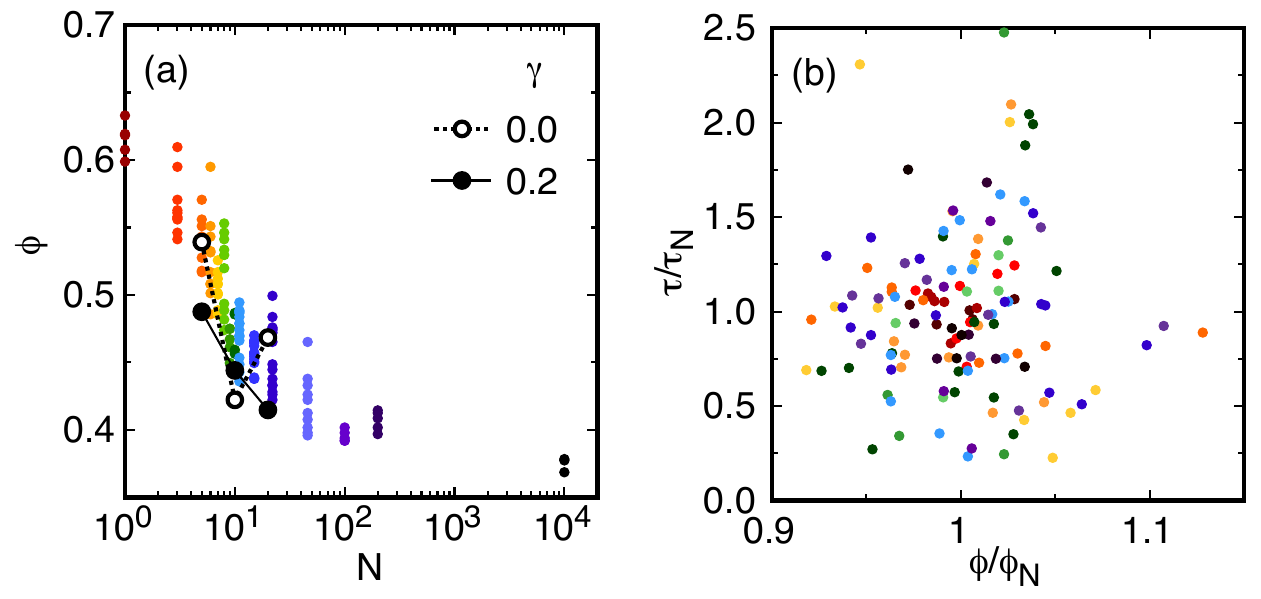}
\caption{(color online) (a) The packing fraction $\phi$ as a function of chain length $N$.  Unconnected circles: repetitions for stress measurements with $\xi=8$ beads at $\gamma=0$ for each chain length $N$. Connected circles:  x-ray tomography measurements with $\xi=7.5$ beads at strain $\gamma=0$ (open circles) and $\gamma=0.2$ (solid circles). Both measurement methods have absolute uncertainties of 0.02.  (b) A scatter plot of the stress $\sigma$ at $\gamma=0.2$ vs.~initial packing fraction $\phi$. The values of $\sigma$ and $\phi$ are normalized by the average values at each $N$ over repetitions at the same $\phi$.   Colors correspond to the same $N$ as in panel a.  The lack of correlation suggests there is not a significant relation between initial packing fraction and strength once the trend with $N$ is removed.
}
\label{fig:phi}
\end{figure}

The global packing fraction $\phi$ is defined for chains as the volume of beads (including the volume inside the beads, but not counting links) divided by the total volume inside the container.  It is measured before the start of triaxial compression ($\gamma=0$) where the bead volume is the number of beads (measured by mass of chains divided by the mass of a single bead and link) times the volume per bead (calculated from the mean bead diameter $a$).  As shown in Fig.~\ref{fig:phi}, for $N=1$ the packing fractions are in the range 0.60-0.63 typical of random packings of spheres \citep{KFLJN94}.  The packing density decreases with increasing chain length as the links reduce freedom of motion which causes chains to pack less densely.  

The global packing fraction was also obtained from the x-ray tomography data at strains $\gamma=0$ and $0.2$ by dividing the sphere volume by the average volume of convex hulls surrounding the beads.  The $N=5$ and $N=20$ samples dilated ($\phi$ decreased) under compression, while the $N=10$ sample was initially loose (presumably due to an inconsistency in packing procedure) so compacted ($\phi$ increased) under compression.  The packing fraction at strains of $\gamma=0.2$ appears to follow a smooth trend with $N$, which is consistent with the expectation that packings  evolve towards a critical packing fraction (a.k.a.~critical state) under shear, regardless of initial packing fraction \citep{SW68}.  The initially loose packing for $N=10$ may have contributed to there being very few links at full extension at $\gamma=0$ (Fig.~\ref{fig:linkdist}b), which made it impossible to resolve the average link strain $\gamma_p$ for $N=10$, and about half the number of entanglements per enclosure as other datasets at $\gamma=0$ (Fig.~\ref{fig:entanglefreeprojareawangle}).  Once the sample was strained to $\gamma=0.2$, these statistics matched more closely with the data with the other strain stiffening packing at $N=20$.

%We note that we observed an overall dilation in experiments that started at high packing fraction rather than compaction.  

To test the effect of initial packing fraction on the packing strength, we show a scatter plot in Fig.~\ref{fig:phi}b of the stress $\sigma$ at $\gamma=0.2$ vs. $\phi$ for 5-10 repetitions at each $N$.  The normalizations $\sigma_N$ and $\phi_N$ are the mean values of $\sigma$ and $\phi$, respectively, average over the repetitions at each $N$.  The scatter suggests there is no significant correlation between initial packing fraction $\phi$ and strength once the effect of chain length $N$ is removed, so that differences in initial $\phi$ in different experiments are not responsible for differences in strength.  %This lack of correlation suggestes that the the large possible variation in initial $\phi$ has a negligible effect on strength.

\section{Calculation of effective free enclosed area $A_{eff}$}
\label{sec:Aeffcalc}

\begin{figure}
\includegraphics[width=3.in]{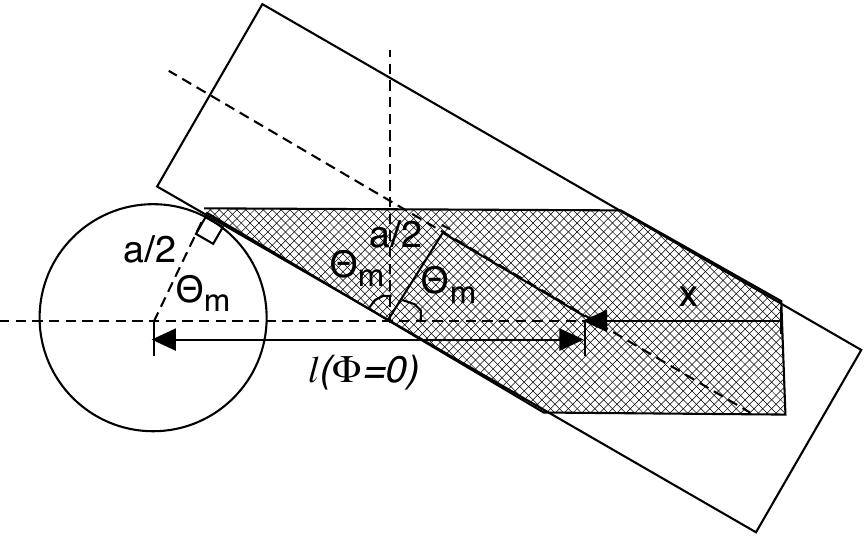}
\caption{Illustration of entangled chain (approximated as a tube) crossing the entanglement manifold (horizontal dotted line).  The beads of the entangling chain limit the positions and angles that an entangled chain can cross the entanglement manifold, producing an excluded volume effect on the fraction of entangled volume filled by chains.  The shaded region shows the cross-section of the volume of the entangled chain inside the entangled volume.
}
\label{fig:Aeffcalc}
\end{figure}

%page 34
The effective free enclosed area $A_{eff}/A = \langle A(\Theta,\Phi)\cos\Theta\rangle$ produced by entangling chains was calculated by averaging over each chain from x-ray tomography data.  The area of the entangling manifold for each chain segment is calculated in $m$ subsegments for each neighboring pair of beads.  The available area in a subsegment of the two-dimensional projection is estimated as the area of the triangle with corners at the centers of the two neighboring beads attached to a link and  the midpoint of the connector as $A= (L_1-a)(L_2-a)\sin\beta$ where $\beta$ is the angle made between the vectors $\vec L_1$ and $\vec L_2$ from the midpoint of the connector to each of the beads.  The subtraction of $a$ approximately accounts for the excluded area where the centers of entangled beads could not fit, treating the chains as cylindrical tubes of diameter $a$.    %The length $a_e$ is a modification of $a$ accounting for the third dimension (into the page of Fig.~\ref{fig:Aeffcalc}) in which a bead of an entangled chain nestled between the two beads bounding the segment of the manifold can have a projected distance into the plane of Fig.~\ref{fig:Aeffcalc} as small as $\sqrt{a^2-\bar d^2/4} = 0.83a$.  An average between this minimum and the maximum of $a$ when passing directly over the beads is $a_e = 0.92$.  $\Theta_m$ should not increase due to $
The factor of $\cos\Theta$ in $A_{eff}$ accounts for the larger entangled volume of a chain at angle $\Theta$ relative to a normal vector to the entanglement manifold, however this slightly overestimates the entangled volume due to cases where the chain extends to $x<0$ if the thickness of the entangled volume $\delta$ is not infinitesimal as seen for example in Fig.~\ref{fig:Aeffcalc} where $\delta =a$.   %A correction factor $f(x,\Theta,\Phi)$ account for the fraction of the entangled chain that is inside the entangled volume for $x>0$.   
The possible orientation angles of the entangled chain up to a maximum $\Theta_m$ are integrated over the distance $x$ from the midpoint of the connector in the plane of the entanglement manifold.  We do not integrate over the perpendicular direction to the entangled manifold as it is expected to be a small correction.  A factor of $\sin\Theta$ is included in the integral for integrating spherical polar coordinates $\Theta$ and $\Phi$, where $\Phi$ is the azimuthal angle around the vertical measured in the entanglement manifold: 
{\small\begin{multline}
\frac{A_{eff}}{A} = \frac{\langle A(\Theta,\Phi)\cos\Theta\rangle}{A} =\\
%\frac{1}{\pi m(L-a)^2} \sum_{m'=1}^{m}\int_0^{L-a}\int_0^{2\pi}\int_0^{\Theta_m} \frac{x\sin\Theta\cos\Theta}{f(x,\Theta,\Phi)}  d\Theta d\Phi dx
\frac{1}{\pi m(L-a)^2} \sum_{m'=1}^{m}\int_0^{L-a}\int_0^{2\pi}\int_0^{\Theta_m} x\sin\Theta\cos\Theta  d\Theta d\Phi dx \ .
\end{multline}}
The  prefactor is such that the integral is 1 in the two-dimensional case which is equivalent to removing the factor $\cos\Theta$ and setting $\Theta_m=\pi/2$.  Integrating over $\Theta$ yields
\begin{equation}
\small\frac{A_{eff}}{A} = \frac{1}{2\pi m(L-a)^2}  \sum_{m'=1}^{m}\int_0^{L-a}\int_0^{2\pi} x(1-\cos^2\Theta_m) d\Phi dx \ .
%\small\frac{A_{eff}}{A} = \frac{1}{2\pi m(L-a)^2}  \sum_{m'=1}^{m}\int_0^{L-a}\int_0^{2\pi} \frac{x(1-\cos^2\Theta_m)}{f(x,\Theta,\Phi)} d\Phi dx
\end{equation}
From Fig.~\ref{fig:Aeffcalc}, $\cos\Theta_m = a/l(\Phi,x)$, where $l(\Phi,x)$ is the distance from the entangling chain at angle $\Phi$ to the point a distance $x$ from the connector in the middle of the segment, resulting in
{\small\begin{multline}
\small\frac{A_{eff}}{A} = \\
\frac{1}{2\pi m(L-a)^2}  \sum_{m'=1}^{m}\int_0^{L-a}\int_0^{2\pi} x\left[1-\frac{a^2}{l(\Phi,x)^2}\right]  d\Phi dx \ .
%\frac{1}{2\pi m(L-a)^2}  \sum_{m'=1}^{m}\int_0^{L-a}\int_0^{2\pi} \frac{x}{f(x,\Theta,\Phi)}\left[1-\frac{a^2}{l(\Phi,x)^2}\right]  d\Phi dx
\label{eqn:Aeffcalc}
\end{multline}}
%For the x-ray data, the integral over $\Phi$ is approximated by  summing over the segments of the entangling chain, where $l(m'',x)$ is the distance from each bead $m''$ to the point at $x$ in the middle of the segment $m'$ according to
%{\small\begin{multline}
%\frac{A_{eff}}{A} = \\
%\frac{1}{m(m+1)(L-a)^2}  \sum_{m'=1}^{m}\sum_{m''=0}^m \int_0^{L-a} x\left[1-\frac{a^2}{l(m'',x)^2}\right]   dx \ .
%\frac{1}{m(m+1)(L-a)^2}  \sum_{m'=1}^{m}\sum_{m''=0}^m \int_0^{L-a} \frac{x}{f(x,m'')}\left[1-\frac{a^2}{l(m'',x)^2}\right]   dx
%\end{multline}}
%correcton for sum over actual phi is 1\% correction on slope, 4\% smaller error
 To approximate the integral over $\Phi$, the value of $\Phi_{m'}$ was calculated for each bead center, and for each value of $\Phi$ of the integral, we calculated $l(\Phi,x)$ based on the bead with the closest $\Phi_{m'}$ or $\Phi_{m'}\pm\pi$.  To account for snaky chains which might come near the entanglement manifold to further exclude volume,  any $x$ value at a point within a diameter $a$ of another bead center of the same subchain of length $m$ was removed from the above calculation so that is has zero contribution to $A_{eff}$ (the term in the integrand is set to zero for that point).  The resulting values calculated from x-ray data are averaged over all contours of all chains, and used for the measured values of $A_{eff}$  for each $m$ in Fig.~\ref{fig:entanglefreeprojareawangle}.

 \begin{figure}
\includegraphics[width=2.8in]{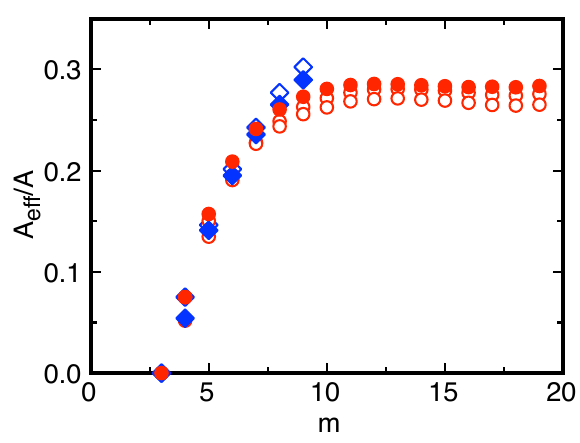}
\caption{The ratio $A_{eff}/A$ accounting for excluded volume effects in three dimensions for chains for different contour lengths $m$. The value is nearly constant in the strong strain-stiffening range ($m\ge 8$), corresponding to the limit of large entangled area widths relative to particle widths used in the entanglement density model of \cite{GFHG12}. 
}
\label{fig:Aeffratio}
\end{figure}

 The ratio of $A_{eff}/A$ is plotted as a function of chain segment length $m$ in Fig.~\ref{fig:Aeffratio}.  We find a consistent $A_{eff}/A = 0.27\pm0.01$ in the range of strong strain stiffening ($m\ge 8$), with the $\pm$  corresponding to the standard deviation at different contour lengths $m$ for both $N=10$ and 20.  This consistency of the ratio $A_{eff}/A$ in this range is in agreement with the earlier expectation that this ratio is a constant for staples \cite{GFHG12}  ($A_{eff}/A$ corresponds to $1/\alpha$ in \cite{GFHG12}).   In the limit of large $m$, the distance from where the entangled chain crosses the entangling manifold to the entangling particle $l(\Phi,x)$ is likely to be large compared to the particle diameter $a$.  In that limit,  the factor $1-a^2/l(\Phi,x)^2$ in the integrand of Eq.~\ref{eqn:Aeffcalc} approaches 1.    For chains, this corresponds to the limit of large $m$ which can have entanglement manifolds much wider than $a$.   However,  for smaller $m \le 8$, the  ratio $A_{eff}/A$ does drop to zero as the constraints of excluded volume become more significant where the entangled width of the entanglement manifold is only slightly larger than $a$ [i.e.~the largest $l(\Phi,x)$ is only slightly larger than $a$].   Using the derivation of Eq.~\ref{eqn:Aeffcalc} as a guide, it can be inferred that the maximum possible value of $A_{eff}/A$ is 1/2 in the extreme limit of thin entangled particles inside a large entangled volume where $a/l(\Phi,x)$ would be small for most of the volume.

%for which $A_{eff}/A \rightarrow 0.5$ from below due to the factor $\cos\Theta$ in the integrand.   is 53\% of the infinite chain limit,
The consistency of $A_{eff}/A$ from $m=8$ to 18 suggests that, like staples \cite{GFHG12},  the number of entanglements could be estimated for a variety of particle shapes, with a single fit parameter $A_{eff}/A = 1/\alpha$ when the width of entangled regions is large compared to the particle thickness.  
%For corrected data, $A_{eff}/A = 0.26\pm0.01$ for $m\ge 8$.

%POSSIBLE CORRECTIONS:
%The fraction of the entangled tube inside inside the entangled volume for $x>0$ is estimated as $f(x)\approx1-(x_c-x)^4/2x_c^4$ where the maximum $x$ where this overcounting of entangled volume happens is $x_c \approx L/2$.   Divide integrand by $f(x,m'')$  makes $A_{eff}$5-20\% larger and a little more linear, but calculation a very rough estimate
%using average distance to chain backbone instead of cylindrical tube: including $a_e=0.92$ reduces $n/A_{eff}$ by 0.1, slight increase in linearity.
%correction for end of chain $1/2N\cos\Theta$->makes $A_{eff}$ bigger, $n/A$ smaller
%removing area when other particles of entangling chain come within $a$ was only correction that caused $A_{eff}$ to become smaller, making n/A_eff bigger
%...correction less than 12\% at each $m$ using $x<0.5(1+\tan\Theta_m)$ as upper bound using orrection faction $[0.5-x/(1+\tan\Theta_m)]$
  %accounting for z-dependence in projected plane only adjusts l(phi) in 3rd decimal place
%These equations overcount the volume of entangled particles when $x<(a/2)(1+\tan\Theta)$ due to chains that extend to $x<0$, which should increase the value of $A_{eff}/A$ by less than 6\% in our case.

\subsection{Monte Carlo calculations of $A_{eff}/A$}

%staples
The calculation of $A_{eff}/A$ for chains shows that the number of entanglements $n$ can be predicted given a particle geometry, but since chains have a flexible shape it required input of x-ray tomography data to get the particle geometry in the packing.  To confirm that such a geometric calculation for $A_{eff}/A$ can make {\em a priori} predictions for rigid particles based on the particle shape without packing information, we carried out Monte Carlo simulations of $A_{eff}/A$ for staples at various aspect ratios of arm length over backbone length.  The geometry of staples is different than Fig.~\ref{fig:Aeffcalc}, assuming cylindrical arms of diameter $a$ in the enclosed rectangular area $A=(L_x-a/2)(L_y-a/2)$.  We randomly sample values of arms crossing the entangled plane at positions $0\le x \le L_x-a/2$, $0 \le y \le L_y/2-a/2$, and orientations $0 \le \Theta \le \pi/2$ (with probability weighted by $\sin\Theta$), and $0 \le \Phi \le \pi/2$.  Valid entanglements with no overlap of particles in the set of coordinates are weighted by a factor of $\cos\Theta$ in the average to account for the larger volume of the staple in the entangling volume, and contribute a factor of zero if there is overlap.  A set of positions and orientations is considered to be a valid entangled position without overlapping another particle  if
\be
x + \left(\frac{1}{\cos\Theta}+\tan\Theta\right)\frac{\cos\Phi}{2} < L_x-a/2
\ee
and
\be
y + \left(\frac{1}{\cos\Theta}+\tan\Theta\right)\frac{\sin\Phi}{2} < L_y-a/2 \ .
\ee
%\begin{equation}
%\frac{A_{eff}}{A} = \frac{1}{2\pi} \int_0^{l-a}\int_0^{l-a}\int_0^{2\pi}\int_0^{\pi/2} \left \frac{1}{\cos\theta}+\ref{1}{\tan\theta}\right]\sin\Theta\cos\Theta d\Theta d\Phi dxdy
%\label{eqn:Aeffcalc}
%\end{equation}
The values of $A_{eff}/A$ obtained from this Monte Carlo calculation for staples are shown in Fig.~\ref{fig:Aeffratiostaples} for different aspect ratios $L_x/L_y$ with a fixed backbone length $L_y = 13a$, corresponding to the simulation of \cite{GFHG12}.

\section{Bending stress calculation}
\label{sec:bendingstresscalc}

 Since most of the bending tends to happen at the corners of staples (Fig.~\ref{fig:staplebends}), this is probably a weak point due to the manufacturing, as the stress would not be expected to be any higher there than in the backbone unless loading points were unusually concentrated around both sides of each corner of each staple. Thus, the stress in the backbone is probably the best estimate where the deformation happens near the corner.  Treating the backbone of the staple as a rectangular beam using standard beam theory, the normal stress in a beam is $\sigma_p=My/I + F_p/wt$, where $M=F_pl$ is the applied torque due to a force $F_p$ at lever arm distance $l$ perpendicular to the backbone, and $y$ is the distance from the neutral axis (center of the backbone).  $I= wt^3/12$ is the area moment of inertia for a rectangular beam.  Since $y=\pm t/2$ on the inner and outer surfaces of the staple backbone, then the average magnitude of normal stress on the inner and outer surface is is $\sigma_{p,s} = 6F_p l/wt^2$.  %The axial stress in the backbone $F_p/wt$ and average shear stress in the arms $F_p/wt$ are expected to be about 4 times smaller for this off-center tensile loading. 
  Since the shear stress in the arms is zero outside of the point of contact, and relatively small in the region between the point of contact and the corner, most of the relevant stress is on the backbone.  While the normal stress is smaller on the thin side of the backbone, this is in part made up for by a larger shear stress near the neutral axis.  Such corrections for the detailed geometry would be much smaller than the fractional uncertainty on $l$.  so the average shear stress on the surface of the particles can be approximated by $\sigma_{p,s} L_y/(L_y+2L_x)$.

\section{Relation between force and plastic deformation of staples}
\label{sec:force_theta}

\begin{figure}
\centerline{\includegraphics[width=2.6in]{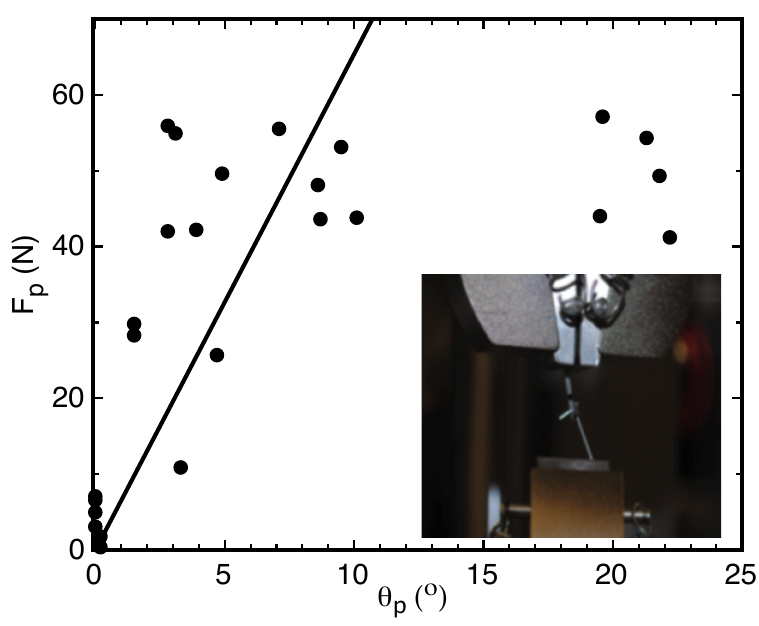}}
\caption{Force $F_p$ required to plastically deform a staple arm to a permanent bend angle $\theta_p$.  The slope $\partial F_p/\partial \theta_p$ is fit to approximate a linear regime of plastic deformation.}
\label{fig:torque_angle}
\end{figure}
%force vs theta -methods
 To measure the slope $\partial F_p/\partial \theta_p$ for plastic deformation of individual pairs of entangled staples, we entangled two staples together and pulled on the opposite ends of the staples to mimic  the arrangement of entangled particles pulling against each other in a random packing.   The pair of  particles were hooked around each other, while clamping the unhooked arms of each staple, as shown in the inset of Fig.~\ref{fig:torque_angle}.  To arrange the particles so that the entangling would provide some resistance to pulling, the clamped arms were rotated 90 degrees in the horizontal plane relative to each other, and the backbones were tilted relative to vertical  so the hook ends would not  slide  apart when initially pulled on.  We then pulled on the staples moving the top clamp  at a rate of 0.05 mm/s  while recording the maximum extensional force $F_p$ applied by the materials tester.  After the peak force, plastic deformation occurred,  typically causing the hooked staples to slip and the force to drop suddenly.   Bend angles $\theta_p$ were measured in terms of a change in angle  between the arms and the backbone relative to the initial 90$^{\circ}$ angle, and this was measured after the force was completely removed to record the angle of permanent plastic deformation.  The reported bend angles $\theta_p$ are the mean %plus or minus the standard deviation of the change 
of the 4 angles -- 2 angles for each staple. 
 %this is recorded as alpha in Henos' measurements
 Each measurement point was done with a new staple to avoid effects of a history of plastic deformation.  We also varied the initial backbone orientations relative to the vertical by 10$^{\circ}$ to 23$^{\circ}$ in different cases in a range where we could obtain stable entangling to provide representative cases for random packings.  In this range, the lever arm distance of the point of contact typically stabilized at $l=0.6\pm0.2$ perpendicular to the backbone due to the curvature near the corner and width of the staples.  While there is a lot of variation in the data from run to run, we  found no clear systematic dependence of the results on this initial orientation.   %assuming that the large slip in 2 staple experiments doesn't occur in packings, because force does not reach a stable value -> particles would slip until they became unhooked and force dropped to zero.  In packing, other constraints must prevent this slipping.
%measured at different rate: $2\times10^{-2}$s$^{-1}$ is much faster speed than stress-strain curve.  could have reduced deformation. 

%force vs theta
A plot of the maximum force applied to each staple arm as a function of the angle of maximum deformation $\theta_p$ is shown in Fig.~\ref{fig:torque_angle}. A linear fit in the range of $\theta_p < 11^{\circ}$ yields the slope $\partial F_p/\partial \theta_p = 6.5\pm0.9$ N$/^{\circ}$, with a 21 N standard deviation of the data around the fit.   
%A constant fit to the plateau for $\theta_p>6^{\circ}$ yields $51$ N with a standard deviation of 5 N  as the ultimate bending force before the staples yield completely.  
The approximation that $\partial F_p/\partial\theta_p$ is a constant oversimplifies the non-linear nature of plastic deformation,  because of the ultimate strength where the load stops increasing with bend angle, and an elastic deformation limit around 10 N before the staples exhibit any permanent plastic deformation.  

\end{document}